\documentclass{article}%[9pt,twocolumn,twoside]

\usepackage{multicol}
\usepackage{graphicx}
\usepackage{color}
\usepackage{amsmath}
\usepackage{amssymb}
\usepackage{enumitem}
\usepackage[margin=0.6in]{geometry}
\usepackage{MnSymbol}
\usepackage{cite}
\usepackage{authblk}
\usepackage[version=4]{mhchem}
\begin{document}

\title{Optical phased arrays for LIDAR: \\beam steering via tunable plasmonic metasurfaces}

\author[1,2,3,$^*$]{Antonino Cal\`a Lesina}
\author[4]{Dominic Goodwill}
\author[4]{Eric Bernier}
\author[1,2]{Lora Ramunno}
\author[1,2,3]{Pierre Berini}

\affil[1]{Department of Physics, University of Ottawa, Ottawa, Canada.}
\affil[2]{Centre for Research in Photonics, University of Ottawa, Ottawa, Canada.}
\affil[3]{School of Electrical Engineering and Computer Science, University of Ottawa, Ottawa, Canada.}
\affil[4]{Huawei Technologies Canada, Ottawa, Canada.}
\affil[$^*$]{antonino.calalesina@uottawa.ca}
\date{}

%%%%%%%%%%%%%%%%%%%%%%%%%%%%%%%%%%%%%%%%%%%%%%%%%%%%%%%%%%%%%
\maketitle
\begin{abstract} 
Controlling the phase and amplitude of light emitted by the elements ({\it i.e.}, pixels) of an optical phased array is of paramount importance to realizing dynamic beam steering for LIDAR applications. In this paper, we propose a plasmonic pixel composed of a metallic nanoantenna covered by a thin oxide layer, and a conductive oxide, {\it e.g.}, ITO, for use in a reflectarray metasurface. By considering voltage biasing of the nanoantenna via metallic connectors, and exploiting the carrier refraction effect in the metal-oxide-semiconductor capacitor in the accumulation and depletion regions, our simulations predict control of the reflection coefficient phase over a range $>330^{\circ}$ with a nearly constant magnitude. We discuss the physical mechanism underlying the optical response, the effect of the connectors, and propose strategies to maximize the magnitude of the reflection coefficient and to achieve dual-band operation. The suitability of our plasmonic pixel design for beam steering in LIDAR is demonstrated via 3D-FDTD simulations. 
\end{abstract}

%We demonstrate control of the reflection coefficient (phase over a phase range $>330^{\circ}$ with a nearly constant magnitude) by voltage biasing the nanoantenna via metallic connectors, and exploiting the carrier refraction effect in the metal-oxide-semiconductor capacitor in the accumulation and depletion regions. 
%The reflection coefficient for the single pixel may not have uniform amplitude and not cover the full 360$^{\circ}$: here we show how this impacts the quality of the steered beam.

\section{Introduction}

%Antonio: Is this phase control?
%Active metasurfaces reviews \cite{Science spatiotemporal metasurfaces paper, Recent Progress in Active Optical Metasurfaces Advanced Optical Materials - Werner}
%Antenna phased arrays are largely used in the microwave and radiowave regimes to shape the radiation pattern. In particular, the real-time reconfigurability of phase and amplitude of each element in the array allows us to realize dynamic beam forming and beam steering.

Most commercial light detection and ranging (LIDAR) scanners are too bulky and slow for applications requiring a high refresh rate, such as self-driving cars and autonomous machines. Optical phased arrays \cite{Heck2017} are much smaller and can tremendously improve the scanning speed as they rely on electronic steering rather than mechanical.
%Most commercial light detection and ranging (LIDAR) systems are too bulky and slow for those applications requiring a high refresh rate due to the fact that the scanning is mechanical. Optical phased arrays \cite{Heck2017} are much smaller, provide a higher scanning speed, and can advance LIDAR technology to enable applications in self-driving cars and autonomous machines. 
An optical phased array is composed of phase-tunable light emitters spatially arranged on a plane, where a single emitter is referred to as a pixel. A pixel can be a source of light, such as a LASER, a waveguide, or a LED. However, a pixel can also be a scatterer, such as a nanostructure in a reflectarray or transmittarray metasurface. 

%; in this paper we will discuss a plasmonic pixel for a reflectarray metasurface. % general, we will refer to an element of the array as a pixel, and in this paper we.  %amplitudes and phases are such that  the shape of the optical beam in the far-field.  and their amplitude and phase tunability is required to steer the optical radiation. a tunable beam steering it required that the light emitters are cntrand is used for light beam steering.  .  In order to achieve this goal, each pixel in the array has to be tunable in such a way that the amplitude and phase of the electric field are controllable. possesses a specific amplitude and phase of the electric field, and the tunability of the pixels in the array is paramount to dynamically shape an optical beam in the far-field. This includes the steering of the optical radiation, which is required %In the optical regime, this corresponds to control the amplitude and phase of the transmitted/reflected/generated light for each single pixel. %In such optical beam steering applications, the ability to dynamically control the phase and the amplitude of the radiation reflected or transmitted by a pixel is required.

Beam steering via optical phased array is achieved by imposing a phase gradient along a certain direction in the array. 
Extraneous beams, {\it e.g.}, produced by grating lobes, need to be minimized in LIDAR applications as they can produce false positives during the scanning process.
%The presence of imperfections in the phase gradient contributes to the formation of .
%LIDAR applications require that grating lobes are as they can produce false positives during the scanning process.  
The pitch $a$ of a matrix of pixels ({\it i.e.}, the spacing between emitters) if too large, can cause grating lobes of similar amplitude as the main lobe. In order to avoid these grating lobes for any steering angle, the pitch should satisfy $a<\lambda/2$, where $\lambda$ is the wavelength in the material in which the array is immersed. This requirement trades off with the necessity of having a large enough pitch so that pixels do not interact with each other (no near-field coupling), and each can be controlled independently. 
Furthermore, if the phase of the field emitted by the pixel does not cover a $360^{\circ}$ range, or its amplitude is not uniform across the phase range, the phase gradient will contain imperfections. These imperfections can be periodic with a period much greater than the pitch, thus producing lobes obeying a diffraction grating-type equation; we call these ``long-period grating lobes''. %These imperfections are due to the fact that the field emitted by the pixel is not ideal, {\it i.e.}, the phase of the field does not cover a $360^{\circ}$ range, and its amplitude is not uniform across the phase range.  %due to the field emitted by the pixel (both phase and amplitude)

Among phase-shifting pixels for optical phased arrays, those based on plasmonic resonances can satisfy $a<\lambda/2$. Thus, plasmonic metasurfaces \cite{Meinzer2014,Genevet2017,Scheuer2017,Chen2018a} offer a platform to implement optical phased arrays that completely avoid grating lobes arising from a too large pitch size. % by allowing a sub-wavelength pitch.
Most plasmonic metasurfaces experimentally demonstrated to date are passive; this means that once fabricated, they cannot be further tuned resulting in a static optical response. Such passive metasurfaces have been proposed for focusing \cite{Papaioannou2018}, achromatic lensing \cite{Chen2018}, perfect absorption \cite{ArroyoHuidobro2017}, colouring \cite{Lee2018b}, performing mathematical operations \cite{Silva2014}, holography \cite{Huang2018}, nonlinear field enhancement \cite{Lesina2015b}, beam structuring in the linear \cite{Yue2016} and nonlinear \cite{Lesina2017} regimes, and biosensing \cite{Zeng2015}. However, there is a growing interest in active plasmonics to control the field emitted by individual nanostructures by tuning their surface plasmon resonances \cite{Jiang2018}, and in tunable metasurfaces with reconfigurable optical response \cite{Kang2019,Shaltout2019} for applications in dynamic holography and optical information encryption \cite{Li2018}, adaptive beamforming in wireless communications \cite{Zhang2018}, chirality switching \cite{Yin2015}, active displays \cite{Duan2017}, polarization conversion \cite{Ratni2017}, metalenses with tunable focus \cite{Wang:19}, dynamic beam steering \cite{Busschaert2019,Forouzmand2018a}, and intensity modulators \cite{Olivieri2015}.

A pixel controllable in phase can be achieved by exploiting, for example, phase-change materials \cite{Zhu2017a}, liquid crystals \cite{Komar2018}, Fermi-level gating in graphene \cite{Sherrott2017}, carrier refraction in semiconductors and transparent conductive oxides (TCOs) including indium-tin oxide (ITO) \cite{Huang2016,KafaieShirmanesh2018}, and the thermo-optic effect in bulk dielectrics \cite{Sun2013}. Phase control has also been demonstrated in arrays of coherently-coupled vertical-cavity surface-emitting lasers \cite{Xun2016}, and via voltage-tunable inter-subband transitions in semiconductor heterostructures exploiting the quantum-confined Stark effect \cite{Lee2014c}. 

Tuning the response of plasmonic nanostructures by exploiting the carrier refraction effect in ITO is promising because of the large change in refractive index available in this material (of the order of unity) \cite{Feigenbaum2010,Lee2014d}. Recent experimental \cite{Huang2016,KafaieShirmanesh2018} and theoretical \cite{Forouzmand2017,Forouzmand2018a} studies of phase control for beam steering and focusing, using metallic or dielectric resonant systems, are encouraging. In \cite{Huang2016,KafaieShirmanesh2018,Forouzmand2017}, out-of-plane field enhancement was exploited in metal-insulator-metal (MIM) nano-resonators, where oxide and ITO were placed between the two metal layers. In these cases, the carrier density variation was induced in ITO via single \cite{Huang2016} or double \cite{KafaieShirmanesh2018} gating. The bottom layer was a plane metallic surface, whereas the other layer was either a linear \cite{Huang2016,Forouzmand2017} or fishbone \cite{KafaieShirmanesh2018} grating. The phase range achieved in MIM configurations was $<300^{\circ}$\cite{Forouzmand2017,Forouzmand2018a} for single gating. To achieve a phase range $>300^{\circ}$ a dual-gated solution was proposed \cite{KafaieShirmanesh2018}. 
%electro-optical modeling \cite{Riedel2017}

Here we propose a plasmonic pixel for phase control which can be used in a reflectarray metasurface for optical beam steering. The pixel is composed of a metallic dipole nanoantenna covered by oxide and ITO. We exploit the perturbation of the permittivity of ITO at locations where the nanostructure creates strong field enhancement in the direction of the nanoantenna axis rather than in the out-of-plane direction as in MIM structures, that is, at the nanoantenna extremities and within its gap. Exploiting the field enhancement in the gap of a dipole nanoantenna to control the phase of its reflection coefficient has yet to be reported - we show how this plays a major role in achieving reflection with a large tunable phase and an approximately constant magnitude of the reflection coefficient. %This reminds what is already done in the microwave regime by producing phase shift via a signal applied to the gap. 

The permittivity of ITO is perturbed via the carrier refraction effect induced in the material, as ITO operates electrically as the semiconductor in a MOS capacitor. In particular, through voltage gating enabled by metallic connectors, we induce a perturbation of the carrier density over a thin ITO region near the nanoantenna via accumulation and depletion processes. In accumulation, the carrier density increases, producing a blue-shift of the epsilon-near-zero wavelength, $\lambda_{ENZ}$, of the ITO in this thin perturbed region. As the carrier density increases further, $\lambda_{ENZ}$ eventually passes through the resonance wavelength of the nanoantenna, and the system transitions from a nanoantenna immersed in a dielectric to a nanoantenna surrounded by a metallic shell. This produces a large red-shift of the nanoantenna resonance wavelength, and as we show in detail, is responsible for the large phase variation observed. We also demonstrate that the phase range can be extended by driving the MOS capacitor into depletion, as depleting the carrier density over an increasing thickness produces an additional sizable phase shift. 

% ($>330^{\circ}$).
%Our design is inspired by microwave phased arrays, where the phase shift is achieved by applying a signal in the gap of the antenna.
%We achieve a large phase variation ($>300^{\circ}$) by inducing, through voltage gating, a 
%with constant amplitude of the reflection coefficient
%in reflectance by perturbing the permittivity of ITO at the locations where the nanostructure creates field enhancement, such as at the extremities and in the gap of the dipole nanoantenna. 
%Summarizing, the transition of $\lambda_{ENZ}$ (red dotted line) through a nanoantenna resonance $\lambda_c$ (black dotted line) by increasing $N$ effects a large variation in the position of the resonance wavelength around $\lambda_c$, which produces the large range of $\angle \Gamma$ plotted in Fig. \ref{fig_g30}(d). 
%In order to exploit a phase shifting pixel in a reflectarray for beam steering, the following four conditions must be satisfied. 
%(1) The pitch (or period) $a$ of a matrix of pixels should satisfy $a\leq\lambda/2$ to avoid secondary lobes due to diffraction (grating lobes), where $\lambda$ is the wavelength in the material where the incident and reflected beams propagate. (2) The pitch should also be large enough so that the pixels do not interact with each other (no near-field coupling), and each can be controlled independently. (3) The phase range should be as close as possible to $360^{\circ}$. (4) And finally, the amplitude of the reflected field should be as uniform as possible over the phase range to minimize the formation of additional lobes. 

Our proposed pixel satisfies the conditions discussed above to minimize grating lobes in optical beam steering, as we will show in Section \ref{sec_pixel} via detailed optical and electrostatic simulations. After presenting the numerical approach (\ref{sec_opt_sim}) and an optimized pixel design via detailed optical simulation (\ref{sec_g30_conn04}), we explain the physical mechanism underlying the large phase shift (\ref{sec_ENZ}), and propose a strategy for increasing the amplitude and uniformity of the reflection amplitude (\ref{sec_g15}). We then explore the effect of the metallic connector position on the resonance of the nanoantenna, and show how this can be exploited to design a dual-band pixel at telecom wavelengths (\ref{sec_conn}). Our ITO perturbation model is justified via detailed electrostatic simulations (\ref{sec_DEVICE}). Finally, the use of our proposed pixel for beam steering is demonstrated in Section \ref{sec_array} via 3D simulations of an array using the finite-difference time-domain method (FDTD).

\section{Plasmonic pixel design}\label{sec_pixel}

\subsection{Optical simulation approach}\label{sec_opt_sim}
%In this section, we present our proposed pixel design with detailed simulations of the optical responses in subsection \ref{sec_g30_conn04}. We explain the physical mechanism underlying the large phase shift in subsection \ref{sec_ENZ}, {\it i.e.}, the translation of $\lambda_{ENZ}$ of ITO through the resonance wavelength of the nanoantenna. In subsection \ref{sec_g15}, we propose a strategy to increase the amplitude of the reflection coefficient and to make it nearly constant. In subsection \ref{sec_conn}, we show the effect of the metallic connector position on the resonance of the nanoantenna, and how this can be exploited to design a dual-band pixel at telecommunications wavelengths. Finally, we provide in subsection \ref{sec_DEVICE} a justification for our ITO perturbation model via detailed electrostatic simulations. 

\subsubsection{Description of the pixel}
\begin{figure}[htbp]
\centering
\includegraphics[width=0.8\textwidth]{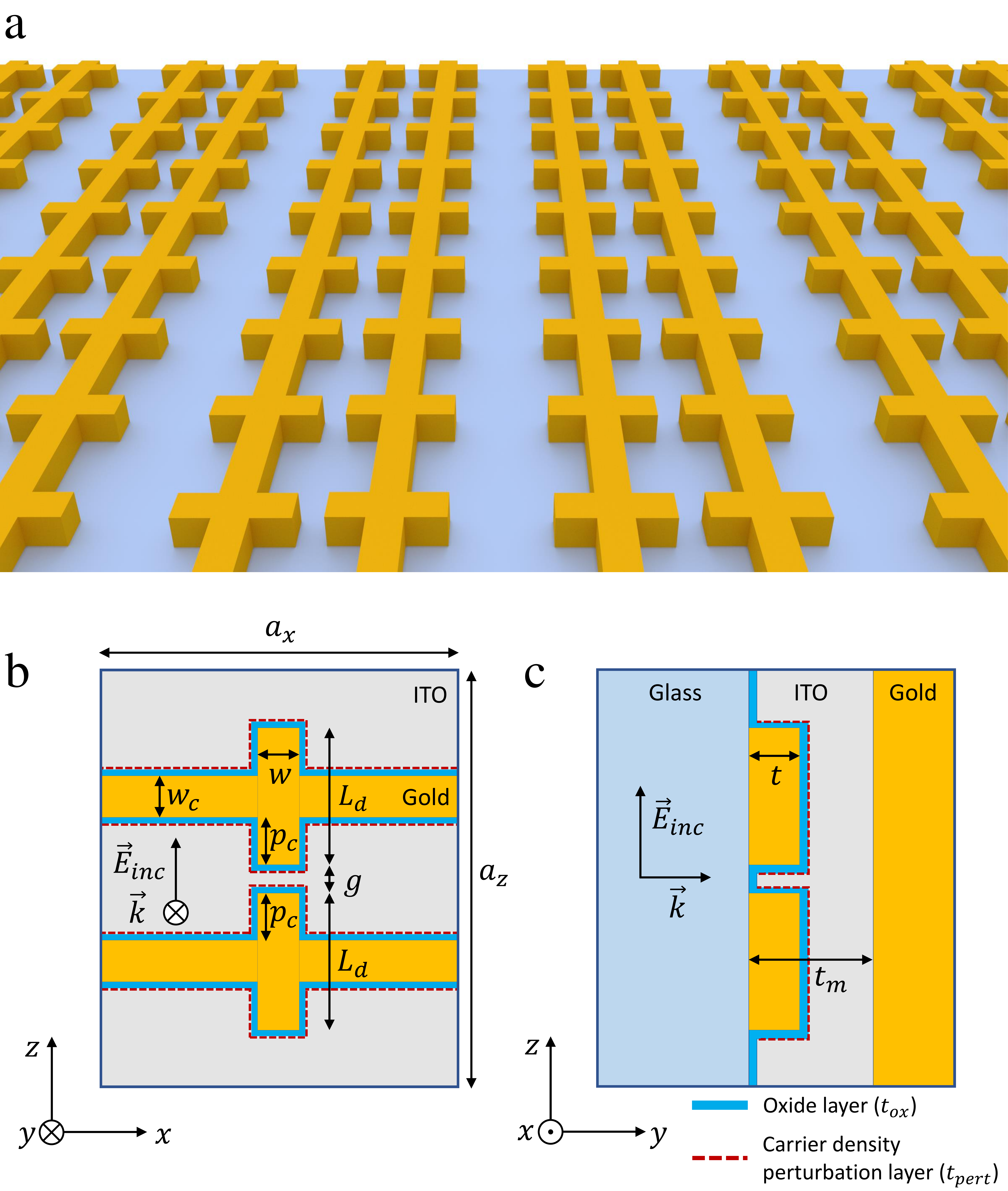}
\caption{(a) Metasurface containing an array of plasmonic pixels (only glass substrate and gold nanoantennas are sketched). (b) Top view, and (c) cross-sectional view of the proposed plasmonic pixel. Throughout this paper, several parameters will remain fixed at the following optimized values: $a_z=500$ nm, $w=50$ nm, $w_c=50$ nm, $t=50$ nm, and $t_{ox}=5$ nm.}
\label{fig1}
\end{figure}

In Fig. \ref{fig1}(a), we show a sketch of a metasurface that uses, as a building block, our proposed plasmonic pixel illustrated in top-view in Fig. \ref{fig1}(b). The pixel has dimensions $a_x$ by $a_z$, and contains a gold dipole nanoantenna, which is formed by two branches of length $L_d$, width $w$, thickness $t$, separated by a gap of size $g$. The pixel also contains two gold lines of width $w_c$, perpendicular to the dipole, which serve as electrical connectors; the distance between the edge of the connector and the edge of the nanoantenna gap is denoted by $p_c$. Here, the connectors are placed in the centre of the branches, so that $p_c=0.5\cdot(L_d-w_c)$, but they can be located at an arbitrary distance $p_c$ from the edge of the nanoantenna gaps.

%they are drawn here at the centre of the branches, but we consider in subsection \ref{sec_conn} how performance changes as a function of connector position $p_c$, which is defined with respect to the edge of the nanoantenna gap.
%The connectors can be located at the center of the branches, {\it i.e.}, $p_c=0.5\cdot(L_d-w_c)$, but can also be shifted along the axis of the nanoantenna. We consider the connectors symmetric with respect to the nanoantenna gap, and the effect of their position will be addressed later on. 
%The nanoantenna and the connectors are assumed to be gold. 
%In Fig. \ref{fig1}(c), we have the side view of the pixel, where we see that 

The nanoantenna of thickness $t$ sits on a glass substrate, and the nanoantenna/glass system is covered conformally by a thin oxide (hafnia) layer of thickness $t_{ox}$, which is then covered by ITO; see the side view of the pixel illustrated in Fig. \ref{fig1}(c). We denote the thickness of the metasurface (metal nanostructure+oxide+ITO) by $t_m$. A gold backplane is placed on top of the ITO allowing us to use the pixel in reflection through the glass substrate. The backplane is depicted here as a flat layer on planarized ITO but it could also be conformal. The electrically-contacted nanoantennas form the metallization of the MOS capacitor and the gold backplane its ground, with the oxide and ITO forming the insulator and semiconductor regions, respectively. In Figs. \ref{fig1}(b) and \ref{fig1}(c), we also indicate by the red dashed line the thin ITO layer of thickness $t_{pert}$, whose refractive index is perturbed by the carrier refraction effect in the MOS capacitor. All corners in our optical computations are square for ease of simulation, as we find that the optical response is not significantly altered by using rounded corners. %(as we show via modelling in subsection \ref{sec_DEVICE}). 

\subsubsection{Simulation details}

The optical modelling was conducted with in-house 3D FDTD software \cite{Taflove2005,Lesina2015}.
Based on fabrication constraints and optical performance evaluated via detailed numerical simulations, we found optimal values for $a_z$, $w$, $w_c$, $t$, and $t_{ox}$ (see the caption of Fig. \ref{fig1}), and they will not be varied throughout this paper. We consider optical (vacuum) wavelengths in the range $\lambda_0=1000$ to 1800 nm due to the availability of high-performance, compact and inexpensive laser sources in that range, and all our designs are optimized to operate at telecom wavelengths, {\it e.g.}, $\lambda_0=1550$ nm.

Gold is modelled using the complex permittivity from \cite{McPeak2015} via the Drude model, {\it i.e.}, $\varepsilon_r(\omega)=\varepsilon_{\infty}-\frac{\omega_p^2}{\omega^2+i\gamma\omega}$, where $\epsilon_{\infty}=8.4156$, $\gamma=4.8257\cdot 10^{13}$ rad/s, and $\omega_p = 1.4117\cdot 10^{16}$ rad/s. %, with fitting parameters valid over the range $\lambda_0=1$--1.7 $\mu$m. 
We choose hafnia (\ce{HfO2}) as our insulator, with a layer thickness $t_{ox}=5$ nm; it is modelled as a lossless and dispersionless dielectric of refractive index $n_{ox}=2.0709$ \cite{Wood1990} (sampled at $\lambda_c=1550$ nm). Glass is simulated as a dispersionless material with $n_{SiO_2}=1.45$.
The permittivity of ITO is modelled via the Drude model with $\varepsilon_{\infty}=4.2345$, $\gamma=1.7588\cdot 10^{14}$ rad/s, and an $\omega_p$ that varies with carrier density $N$ according to $\omega_p=\sqrt{Ne^2/(\varepsilon_0 m_n^*)}$, where $e$ is the electron charge, $\varepsilon_0$ the vacuum permittivity, and $m_n^*=0.35\cdot m_e$ the effective mass of electrons; $m_e$ is the free electron mass. The unperturbed carrier density of ITO is taken as $N_0=3\cdot 10^{20}$ cm$^{-3}$, which corresponds to a plasma frequency $\omega_p=1.652\cdot 10^{15}$ rad/s \cite{Huang2016,KafaieShirmanesh2018}. 

To optimize the design, we consider an infinite 2D array of the plasmonic pixel, though we simulate only a single unit cell containing one pixel.  The unit cell is excited with a $z$-polarized broadband plane wave pulse propagating along $y$, and periodic boundary conditions are applied along $x$ and $z$. We discretize all space dimensions with a uniform space step of 1 nm \cite{Lesina2015}. We calculate the reflection coefficient $\Gamma(\omega)=E_z(\omega)/E_{inc}(\omega)$, from the known incident field $E_{inc}(\omega)$ and reflected field $E_z(\omega)$; the latter is the only nonzero far-field component due to the $z$-polarized applied excitation. The reflected field is obtained in the scattered field region at a distance of 600 nm from the nanoantenna, which can be considered far-field for plasmonic systems. At this distance, the reflected field is a plane wave due to the sub-wavelength pixel size and consequent lack of diffraction orders ($a_x$ and $a_z<\lambda/2$, where $\lambda=\lambda_0/n_{\ce{SiO2}}$, and $\lambda_0$ is the vacuum wavelength). The reflectance is defined as $|\Gamma|^2$ and represents the attenuation factor applicable to the optical power.

\subsubsection{Carrier refraction effect}
From the Drude model introduced above, we note that the plasma frequency scales with $\sqrt{N}$: this is the origin of the carrier refraction effect. 
The carrier density can be varied within a thin layer of ITO by exploiting the operation of a MOS capacitor, where the metal nanoantenna represents the gate (connected to a voltage $V_g$), ITO is used as the semiconductor (connected to a voltage $V_s=0$ V or ground), and a thin insulating (oxide) layer is placed in between. We consider both connectors of the pixel pinned to the same potential (symmetric perturbation), but dual-gating can also be exploited. As explored in more detail via electrostatic simulations in subsection \ref{sec_DEVICE}, when a voltage bias $V_g$ is applied to the gate relative to ground, the carrier density changes within a thin ITO layer due to accumulation ($V_g>V_{fb}$) or depletion ($V_g<V_{fb}$) processes, where $V_{fb}$ is the flat band voltage of our MOS capacitor. 
This local change in carrier density produces a local change in permittivity due to the carrier refraction effect. Since the ITO layer with perturbed carrier density is located in the enhanced field region near the plasmonic nanostructure, its refractive index variation modifies the resonance condition of the nanoantenna and, therewith, the phase of the reflected field. 

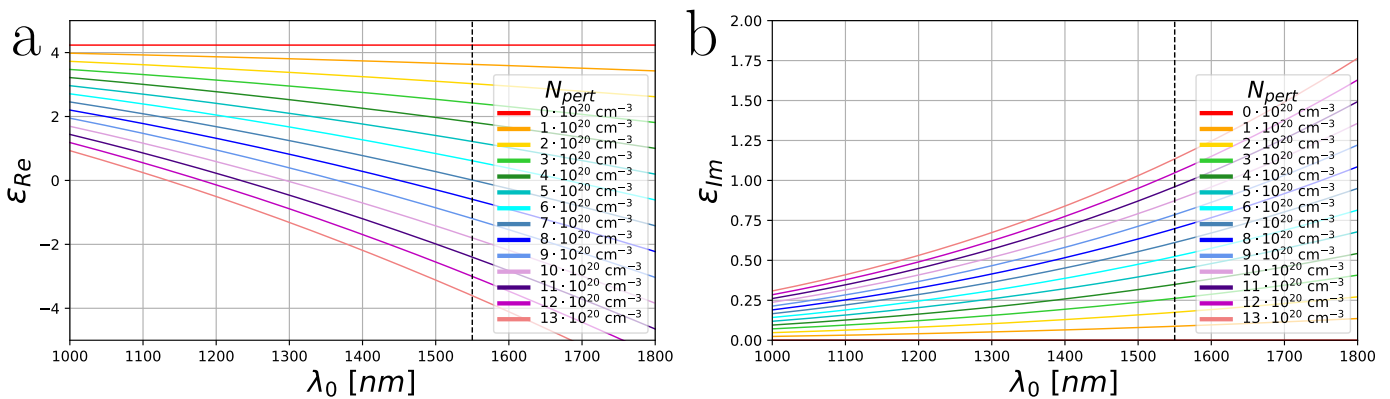
\begin{figure}[htbp]
\centering
\includegraphics[width=1\textwidth]{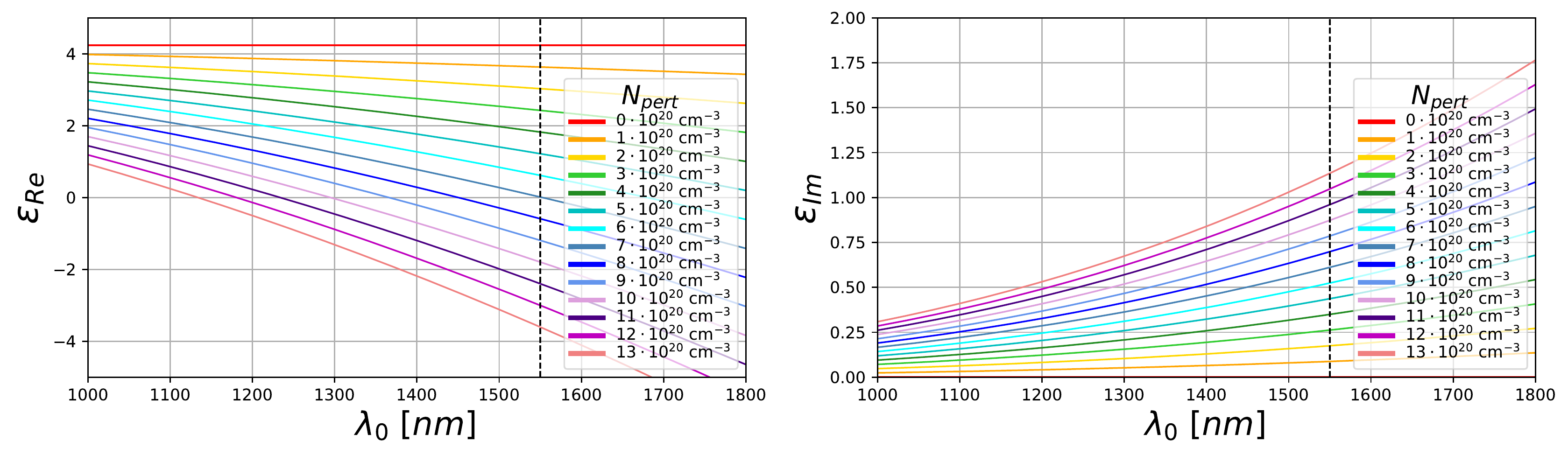} %_Lm-110_Ox3_x290_z500
\caption{(a) $\varepsilon_{Re}$ and (b) $\varepsilon_{Im}$ of ITO {\em vs}. $\lambda_0$ for varying $N_{pert}$. This permittivity is applied within the perturbed ITO layer of thickness $t_{pert}$ in the optical model.}
\begin{picture}(0,0)
\put(-255,180){\Huge a}
\put(0,180){\Huge b}
\end{picture}
\label{fig4}
\end{figure}

Based on the electrostatic analysis of the MOS capacitor in subsection \ref{sec_DEVICE}, we model the local carrier refraction effect in ITO by introducing in the optical model a thin layer of ITO with thickness $t_{pert}=1$ nm, as shown in Figs. \ref{fig1}(b) and (c). The carrier density within the $t_{pert}$ layer is considered uniform and denoted by $N_{pert}$. The permittivity of the $t_{pert}$ layer is shown in Fig. \ref{fig4}, and we see that the epsilon near zero wavelength $\lambda_{ENZ}$ -- that is, the wavelength for which $\varepsilon_{Re}=0$ -- blue shifts with increasing $N_{pert}$. Here $N_{pert}$ is allowed to range between 0 and $N_0$ in depletion, and $N_0$ and $13\cdot10^{20}$ cm$^{-3}$ in accumulation, for a $V_g$ constrained within the breakdown voltage limits, {\it i.e.}, $|V_g|<V_{bk}=3.2$ V. Under negative bias, as we will show in subsection \ref{sec_DEVICE}, the width of the depletion region increases up to $t_{pert}=2$ nm for $V_g=-V_{bk}$, and this case is considered in optical simulations as well. %as we will show in subsection \ref{sec_DEVICE}. 

%When no voltage is applied ($V_g=0$ V), the MOS capacitor is slightly in depletion; we need to apply $V_g=V_{fb}$ is order to have $N_{pert}=N_0$ within the $t_{pert}$ layer. 
%Considering the results in subsection \ref{sec_DEVICE}, it is reasonable to assume in our optical simulations $t_{pert}=1$ nm, and a carrier density $N_{pert}$ ranging  
%Based on the electrostatic analysis of the MOS structure, we model the local carrier refraction effect in ITO by introducing in the optical model a thin layer of ITO with thickness $t_{pert}=1$ where the carrier density is perturbed, The.
 %As we show below, this carrier density variation is localized to a thin layer of semiconductor of the order of $\sim 1$ nm. 

\subsection{Optical performance of plasmonic pixel}\label{sec_g30_conn04}

%We consider the evolution of the reflection coefficient as $N_{pert}$ changes within the perturbed ITO layer as a result of voltage biasing. 
%the carrier density in ITO is perturbed through a thin ITO layer $t_{pert}=1$ nm around the dipole nanoantenna with carrier density $N_{pert}$. As $N_{pert}$ increases, the complex permittivity of ITO varies as shown in Fig. \ref{fig4}.
%Recalling the effects of the connector position (Fig. \ref{fig5}), 
%with parameters $L_d=194$ nm, $g=30$ nm, $a_x=500$ nm, $t_m=89$ nm, and connectors located at a position $p_c=0.4\cdot (L_d-w_c)$ with respect to the edge of the nanoantenna gap.
%such that the structure exhibits only one resonance at the same location where the resonance would be without the connectors (the connectors located at this position are optically non-invasive). 
%By increasing $N_{pert}$, $\lambda_{ENZ}$ blue shifts, as shown in Fig. \ref{fig4}(a), and this causes a large phase variation in the reflection coefficient, as discussed later. %We also validated that the perturbation along $z$ plays the major role for the phase shift mechanism.
%This is very important and, as discussed later, is responsible for the large phase range of the reflection coefficient. 

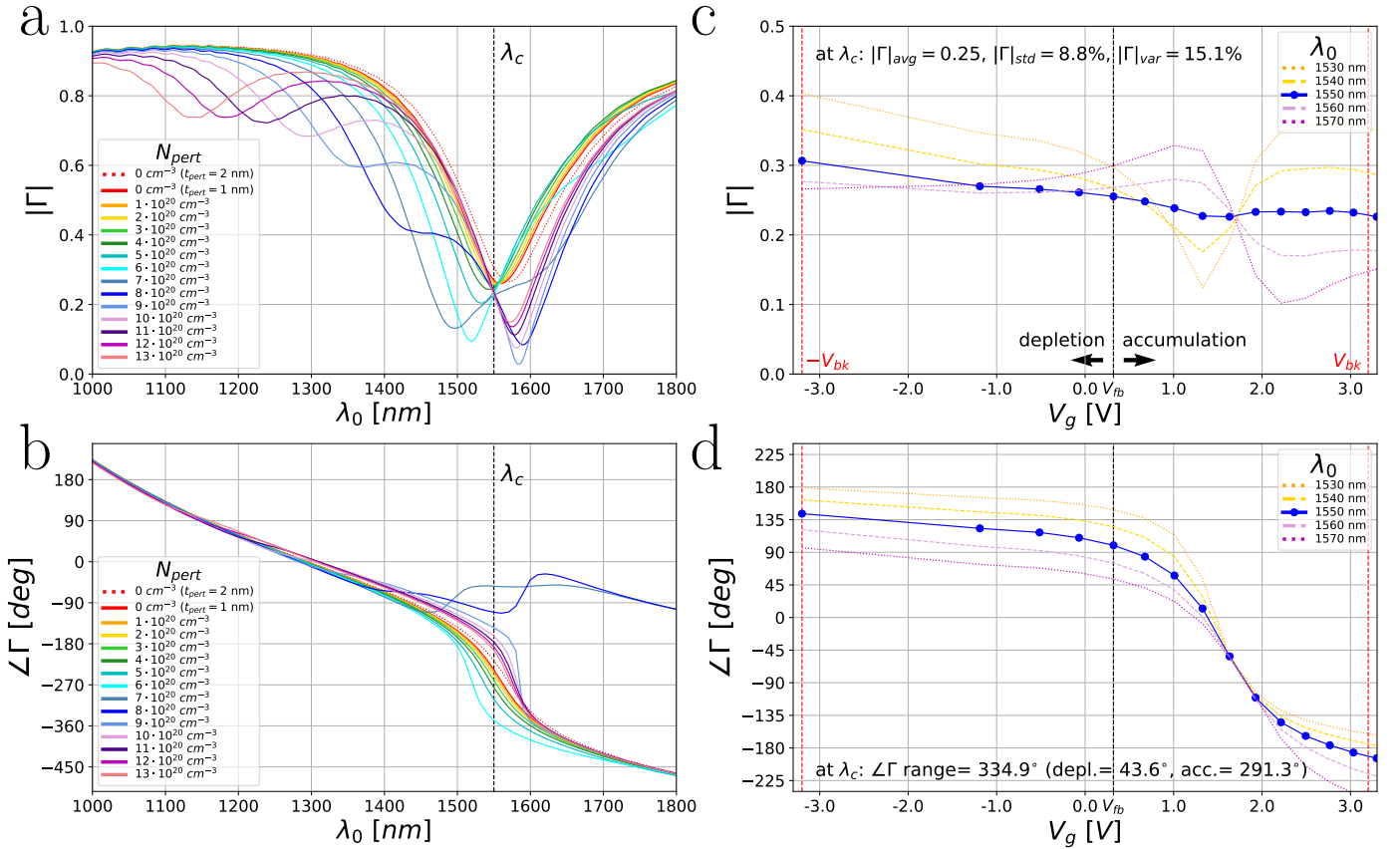
\begin{figure}[htbp]
\centering
\includegraphics[width=1\textwidth]{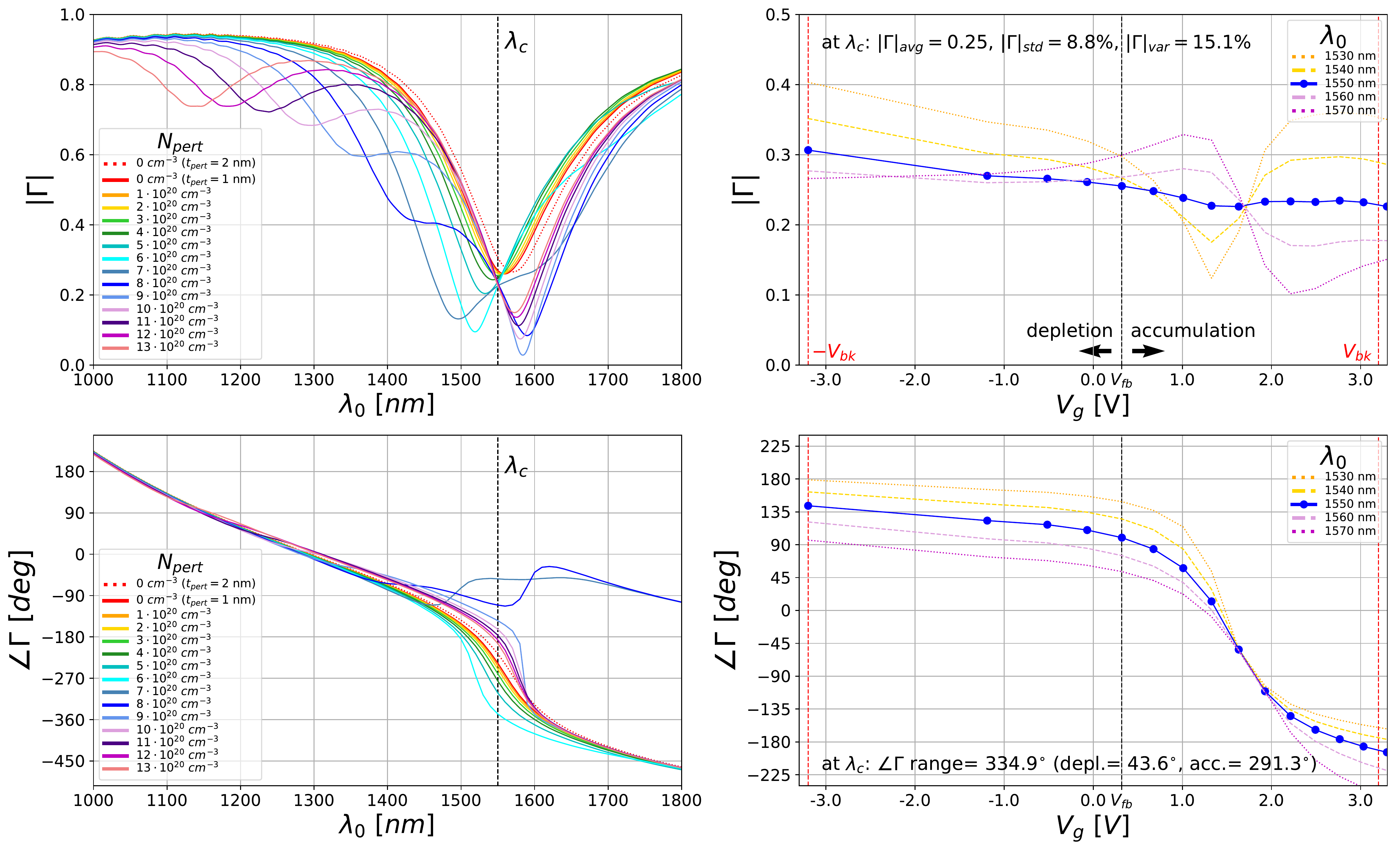} 
\caption{An optimized pixel with parameters $g=30$ nm, $L_d=194$ nm, $a_x=500$ nm, $t_m=89$ nm, and $p_c=0.4\cdot (L_d-w_c)$: (a) $|\Gamma|$ and (b) $\angle \Gamma$ {\it vs} $\lambda_0$ for varying $N_{pert}$ within the $t_{pert}$ layer. (c) $|\Gamma|$ and (d) $\angle \Gamma$ {\it vs} $V_g$ for $\lambda_0$ near $\lambda_c$. The phase plots are unwrapped for clarity.}
\begin{picture}(0,0)
\put(-255,370){\Huge a}
\put(0,370){\Huge c}
\put(-255,205){\Huge b}
\put(0,205){\Huge d}
\end{picture}
\label{fig_g30_conn04}
\end{figure}

Now that we have introduced the concept of our plasmonic pixel and how it operates, we turn to the calculation of the reflection coefficient and how it changes as a function of $N_{pert}$ (and thus, gate voltage) in the perturbed ITO layer. 
We show in Fig. \ref{fig_g30_conn04} the results for an optimized pixel (with dimensions as reported in the figure caption). 
In Fig. \ref{fig_g30_conn04}(a), we plot the absolute value of the reflection coefficient $|\Gamma|$ as a function of $\lambda_0$ and $N_{pert}$ (15 simulation entries). 
The most obvious and remarkable feature of Fig. \ref{fig_g30_conn04}(a) is that there is a wavelength -- called the constant amplitude wavelength, $\lambda_c$ -- for which $|\Gamma|$ has approximately the same value for all values of $N_{pert}$ -- we call this the constant amplitude point, corresponding to a set of 15 reflection coefficient amplitudes, that are almost coincident. The optimization of the pixel is conducted such that $\lambda_c$ coincides with the operating wavelength, {\it i.e.}, $\lambda_c=1550$ nm (vertical black dashed line). 
Furthermore, plotting the phase of $\Gamma$ ($\angle\Gamma$) in Fig. \ref{fig_g30_conn04}(b), we see that at $\lambda_c$ (vertical black dashed line) there is a large variation with $N_{pert}$.

%there is a wavelength $\lambda_c$ -- for this design we have $\lambda_c = 1550$ nm -- for which we can optimize the design so that at the operating wavelength $\lambda_0=1550$ nm
%We find that there exists a wavelength $\lambda_c$ (vertical black dashed line) for which the amplitude of the reflection coefficient is nearly constant for all values of $N_{pert}$ -- we call this the constant reflection amplitude point, $|\Gamma|_c$. Further, we find that at $\lambda_c$, the reflection coefficient phase varies over a range $>300^{\circ}$ as $N_{pert}$ is varied from 0 to $13\cdot10^{20}$ cm$^{-3}$. These two features in the optical response allow for the realization of beam steering at $\lambda_c$.
%In the $|\Gamma|$ curves plotted in Fig. \ref{fig_g30_conn04}(a), we also note that there is a resonance dip which blue-shifts for increasing $N_{pert}$; this dip is located close to the epsilon near-zero wavelength of the perturbed ITO layer for increasing $N_{pert}$ (see Fig. \ref{fig4}(a)), and its role in shifting the phase of the reflection coefficient will be discussed in the next subsection.
%all curves cross at a constant amplitude wavelength $\lambda_c= 1550$ nm (black dashed line). One characteristic of $\lambda_c$ is that $|\Gamma|$ remains approximately constant with $N_{pert}$. The other characteristic becomes evident from the phase of $\Gamma$, $\angle\Gamma$, plotted in Fig. \ref{fig_g30_conn04}(b), where we notice a large variation with $N_{pert}$ at $\lambda_c$. 

Using the relation between $V_g$ and $N_{pert}$ obtained by electrostatic simulations in subsection \ref{sec_DEVICE}, we plot $|\Gamma|$ and $\angle \Gamma$ in Figs. \ref{fig_g30_conn04}(c) and \ref{fig_g30_conn04}(d), respectively, as functions of $V_g$ for wavelengths close to $\lambda_c$. %The set of $|\Gamma|$ values at $\lambda_c$ for the 15 simulation cases can also be referred as $|\Gamma|_c$ from now on, and for this set we will provide statistics.
Here it is evident that at $\lambda_c$, $|\Gamma|$ is nearly constant and $\angle \Gamma$ goes through a large variation. These are desirable characteristics for the realization of beam steering at $\lambda_c$ with long-period grating lobes within an acceptable level.  
Considering all the 15 simulation entries (highlighted as blue dots in Fig. \ref{fig_g30_conn04}(c)) from Fig. \ref{fig_g30_conn04}(a) at $\lambda_c$, we obtain an average reflection coefficient $|\Gamma|_{avg}\sim 0.25$.
Not only should $|\Gamma|_{avg}$ be large, but the sequence of blue dots in Fig. \ref{fig_g30_conn04}(c) should also be flat. This means that the percentage standard deviation and the percentage variation of $|\Gamma|$, indicated as $|\Gamma|_{std}$ and $|\Gamma|_{var}=\frac{|\Gamma|_{max}-|\Gamma|_{min}}{|\Gamma|_{max}+|\Gamma|_{min}}$, respectively, and reported in Fig. \ref{fig_g30_conn04}(c), have to be as small as possible, ideally zero, in order to reduce the level of the long-period grating lobes of a steered beam. 

%the percentage variation $|\Gamma|_{var}$ . 
%a percentage standard deviation of $|\Gamma|$ of $|\Gamma|_{std}\sim 9\%$, and a percentage variation of $|\Gamma|$ of $|\Gamma|_{var}=\frac{|\Gamma|_{max}-|\Gamma|_{min}}{|\Gamma|_{max}+|\Gamma|_{min}}\sim 15\%$, as reported in Fig. \ref{fig_g30_conn04}(c). These statistical parameters are 

As reported in Fig. \ref{fig_g30_conn04}(d), the phase range achieved at $\lambda_c$ within the voltage breakdown limit is $\sim334^{\circ}$, where $\sim 291^{\circ}$ is obtained in accumulation and $\sim 43^{\circ}$ in depletion. In accumulation, $\lambda_{ENZ}$ of the perturbed ITO layer passes through $\lambda_c$ as $N_{pert}$ increases; this is evident in Fig. \ref{fig_g30_conn04}(a), where the $|\Gamma|$ curves show a resonance dip close to $\lambda_{ENZ}$ which blue-shifts for increasing $N_{pert}$. This mechanism plays the main role in shifting the phase of the reflection coefficient, as we explain in more detail in subsection \ref{sec_ENZ}. As already mentioned, we also consider the case $N_{pert}=0$ cm$^{-3}$ with $t_{pert}=2$ nm, which can be reached by the MOS capacitor in depletion at the limit of the oxide breakdown. This simulation condition is reported in Fig. \ref{fig_g30_conn04}(a) as the red dotted curve, and shows a red shift of the response with respect to $\lambda_c$. The red shift is due to the larger $t_{pert}$, and results in an increase in $|\Gamma|$ in Fig. \ref{fig_g30_conn04}(c) as $V_g$ approaches $-V_{bk} = -3.2$ V.

%Under negative bias, the width of the depletion region over which the refractive index is perturbed increases up to $t_{pert}=2$ nm for $V_g=-V_{bk}$ as we will show in subsection \ref{sec_DEVICE}. 
% due to non ideal phase range and amplitude of the field emitted by the pixel. %In Fig. \ref{fig_g30_conn04}(c), we summarize the properties of $|\Gamma|$ at $\lambda_c$  by calculating mean value $|\Gamma|_{avg}$, percentage standard deviation $|\Gamma|_{std}$, and percentage variation $|\Gamma|_{var}=\frac{|\Gamma|_{max}-|\Gamma|_{min}}{|\Gamma|_{max}+|\Gamma|_{min}}$ for all simulation entries (see blue dots) obtained from Fig. \ref{fig_g30_conn04}(a).
%The large phase range achieved at the crossing-point can be understood by considering the evolution of the system with increasing $N_{pert}$. 
%Fig. \ref{fig_g30}(a) shows that the system produces two resonances in the unperturbed case (see curve for $N=3\cdot 10^{20}cm^{-3}$): one resonance is due to the monopole nanoantenna and one resonance coincides with the ENZ wavelength $\lambda_{ENZ}$. 
%In Fig. \ref{fig_g30_conn04}(a), the curve for $N=3\cdot 10^{20}$ cm$^{-3}$ corresponds to the unperturbed ITO ($N=N_0$) %, flatband) already discussed in relation to Fig. \ref{fig5}. 
%By increasing $N_{pert}$, the plasma frequency increases and the associated plasma wavelength decreases. This means that $\lambda_{ENZ}$ for the perturbed ITO layer decreases with increasing $N_{pert}$ as shown in Fig. \ref{fig4}(a) and sweeps through the resonance of the antenna. 

\subsection{Effect on phase shift due to changing $\lambda_{ENZ}$}\label{sec_ENZ}

\begin{figure}[htbp]
\centering
\includegraphics[width=1\textwidth]{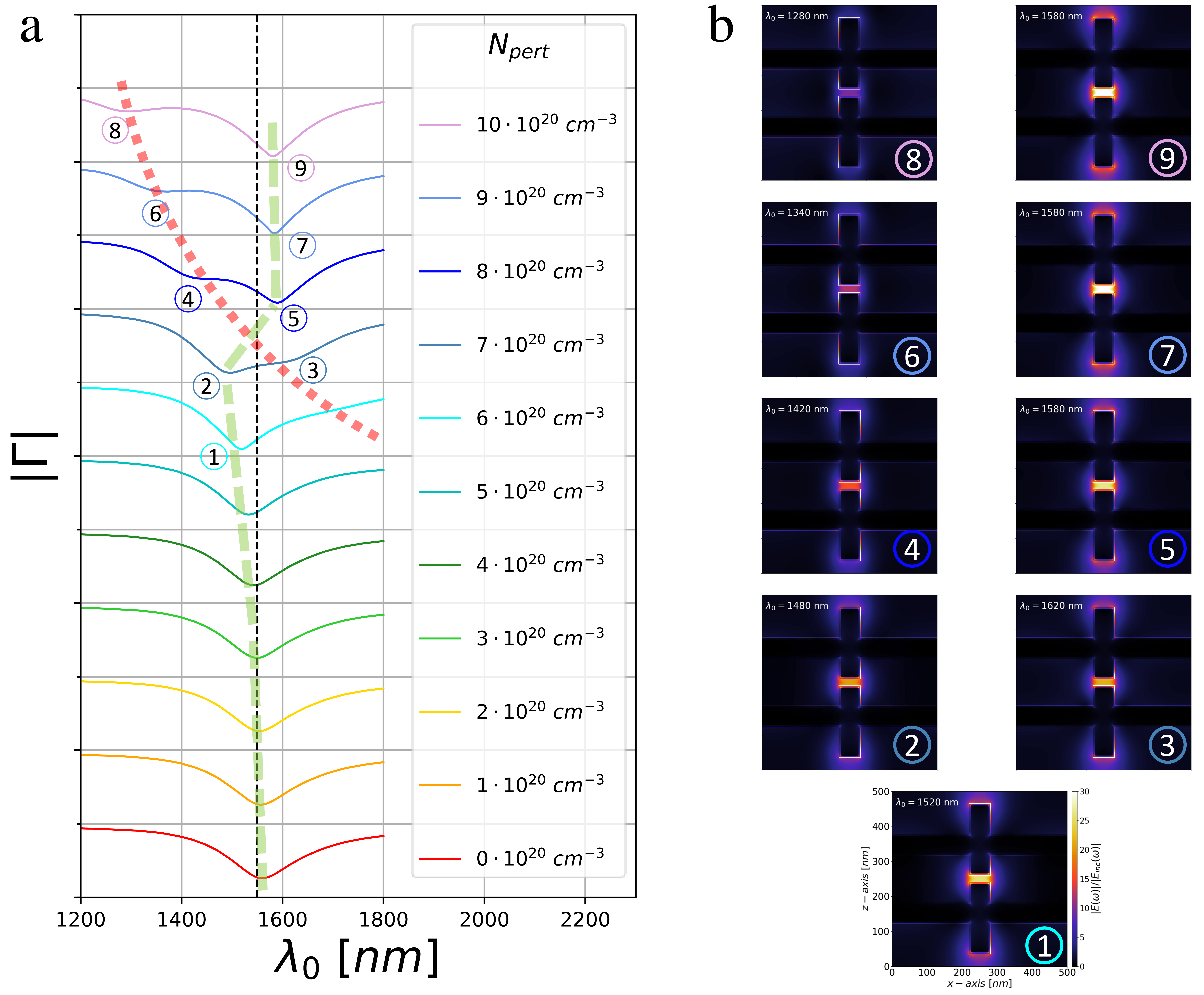} %_Ld-160_Ox3_x326_z500
\caption{Response of the optimized pixel of Fig. \ref{fig_g30_conn04}: (a) $|\Gamma|$ {\it vs} $\lambda_0$ for varying $N_{pert}$ as indicated in the legend. (b) Field distribution within the pixel for wavelengths corresponding to the resonances identified by numerals in Fig. \ref{conn04_unwrapped}(a).}
\label{conn04_unwrapped}
\end{figure}

To better understand the operation of our plasmonic pixel and see more clearly how the resonance dips evolve with increasing $N_{pert}$, we plot in Fig. \ref{conn04_unwrapped}(a) the spectra from Fig. \ref{fig_g30_conn04}(a), but spread out along the horizontal axis. The operating wavelength $\lambda_c=1550$ nm is indicated by the vertical black dashed line. As we can see, this wavelength coincides with the resonance dip of the system in the unperturbed case ($N_{pert}=3\cdot 10^{20}$ cm$^{-3}$). The red dotted curve in Fig. \ref{conn04_unwrapped}(a) tracks the value $\lambda_{ENZ}$ within the perturbed ITO layer as $N_{pert}$ is varied, which we can see is associated with a closely located resonance dip. The green dashed line tracks the evolution of the main nanoantenna resonance dip. 

When $\lambda_{ENZ}>\lambda_{c}$, the perturbed ITO layer around the nanoantenna is dielectric and its refractive index decreases with increasing $N_{pert}$, producing a blue shift of the main nanoantenna resonance dip (see green dashed line for $N_{pert}\leq 7\cdot 10^{20}$ cm$^{-3}$). This suddenly changes when $\lambda_{ENZ}$ passes through $\lambda_{c}$ (see intersection of red dotted curve and black dashed line). For $\lambda_{ENZ}<\lambda_{c}$, the perturbed ITO layer becomes a metallic sheath, which makes the nanoantenna effectively larger and produces a sudden red-shift of the resonance wavelength (see jump of the green dashed line for $N_{pert}> 7\cdot 10^{20}$ cm$^{-3}$). 

% and the evolution from resonance 2 to 5 in Fig. \ref{conn04_unwrapped}(a)). 
%This phase shift mechanism is mainly determined by the perturbed ITO surfaces which are  along the $z$-axis, {\it i.e.}, where we have the the overlap between field enhancement and perturbed ITO.
%The nanoantenna evolves from being immersed in a dielectric
% which changes the resonance condition of the nanoantenna. 
%(green dotted line for $N_{pert}\leq 7\cdot 10^{20}$ cm$^{-3}$), (green dotted line for $N_{pert}\geq 8\cdot 10^{20}$ cm$^{-3}$) -- see evolution from resonance 1 to 5 in Fig. \ref{conn04_unwrapped}(a). %We point out that a larger $t_{ox}$ would enhance this effect, causing a larger red-shift and increasing the phase range accessible by the antenna (a larger $t_{ox}$ effectively lengthens the antenna when surrounded by the sheath). %, and we see from the Drude model that $\lambda_{ENZ}$ blue shifts with increasing $\omega_p$, {\it i.e.}, with increasing $N$.

We need to point out that this phase shift mechanism can only work if $\lambda_{ENZ}$ for the unperturbed ITO is greater than the resonance wavelength of the nanoantenna. This condition can be satisfied in ITO and other TCO materials, where the free carrier density can be regulated so that $\lambda_{ENZ}$ is located in the infrared. 

In Fig. \ref{conn04_unwrapped}(b), we plot the field distributions associated with the dips labelled by the encircled numerals in Fig. \ref{conn04_unwrapped}(a).
We observe the evolution of the resonance mode (tracked by the green dashed line in Fig. \ref{conn04_unwrapped}(a)), and find that it corresponds to a notable field enhancement in the gap of the dipole nanoantenna. This mode evolves following the field distributions labelled by 1-2-5-7-9 in Fig. \ref{conn04_unwrapped}(b) and this series contains the  $\lambda_{ENZ}$ jump across $\lambda_c$.
The mode associated with the evolution of $\lambda_{ENZ}$, indicated by the red dotted curve in Fig. \ref{conn04_unwrapped}(a), follows the field distributions labelled by 3-4-6-8  in Fig. \ref{conn04_unwrapped}(b). This mode is characterized by field enhancement in the thin perturbed ITO layer, as is particularly evident in field distributions 4, 6 and 8 of Fig. \ref{conn04_unwrapped}(b). We notice that for $N_{pert}= 7\cdot 10^{20}$ cm$^{-3}$, $\lambda_{ENZ}\sim\lambda_c$, the resonance mode of the nanostructure is hybridized, and the field distributions labelled 2 and 3 in Fig. \ref{conn04_unwrapped}(b) are similar.

\subsection{Optimizing the reflection coefficient amplitude and uniformity}\label{sec_g15}

\begin{figure}[htbp]
\centering
\includegraphics[width=1\textwidth]{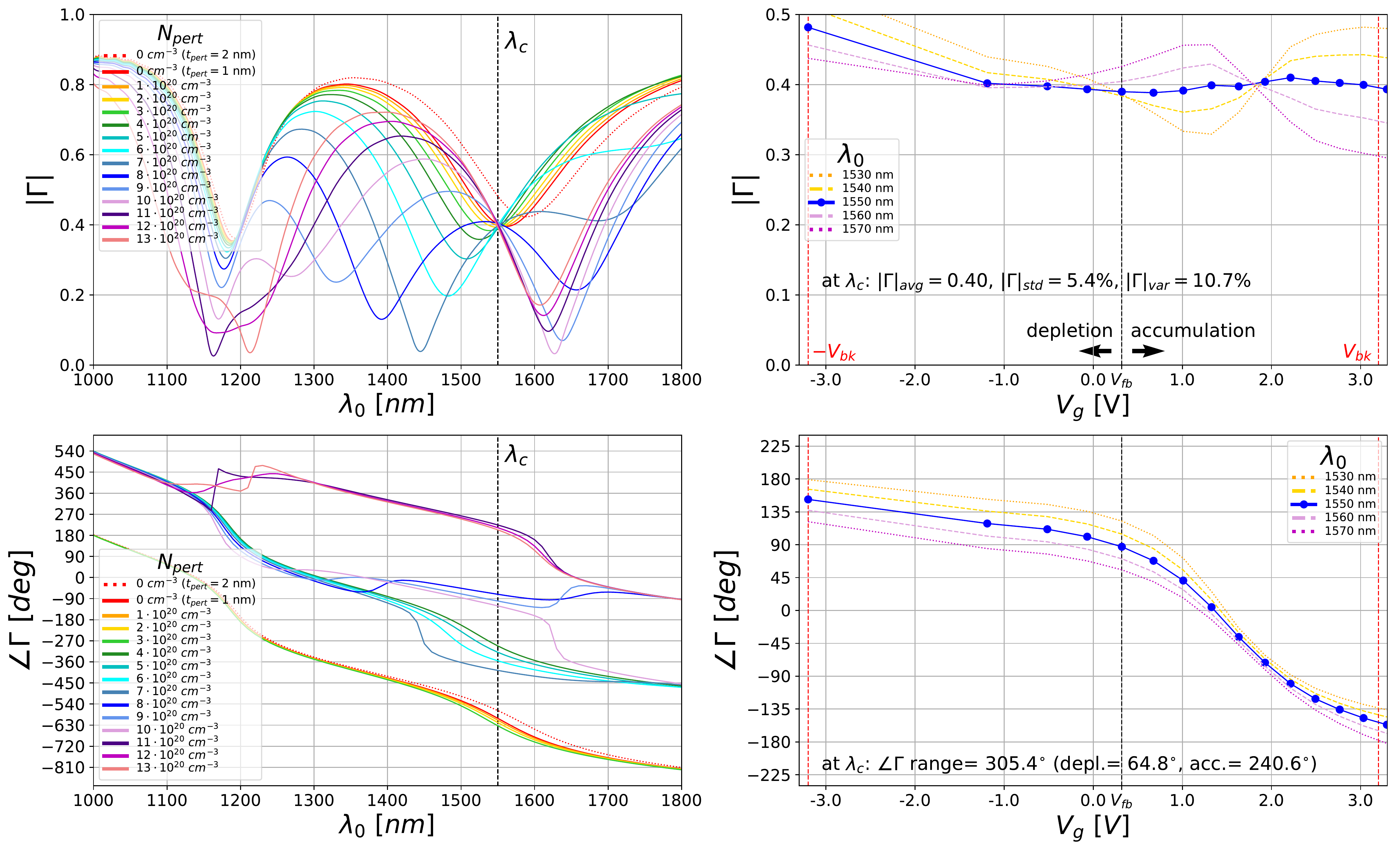} 
\caption{Second optimized pixel with parameters $g=15$ nm, $L_d=149$ nm, $a_x=298$ nm, $t_m=100$ nm, and $p_c=0.5\cdot (L_d-w_c)$: (a) $|\Gamma|$ and (b) $\angle \Gamma$ {\it vs} $\lambda_0$ for varying $N_{pert}$ within the $t_{pert}$ layer. (c) $|\Gamma|$ and (d) $\angle \Gamma$ {\it vs} $V_g$ for $\lambda_0$ near $\lambda_c$. The phase plots are unwrapped for clarity.}
\begin{picture}(0,0)
\put(-255,370){\Huge a}
\put(0,370){\Huge c}
\put(-255,205){\Huge b}
\put(0,205){\Huge d}
\end{picture}
\label{fig_g15}
\end{figure}

In this subsection we describe strategies for increasing the reflectance at $\lambda_c$ in the interests of increasing the power efficiency in beam steering applications, and making the reflectance as uniform as possible with varying $V_g$. To increase pixel reflectance, we increase the field enhancement in the gap of the dipole nanoantenna by reducing the gap length. As the gap must contain two oxide layers, this sets a limit on the smallest possible gap. In Fig. \ref{fig_g15}, we report the optical simulation results for a second pixel optimized to exhibit a constant amplitude point at $\lambda_c$. In this case, the average reflection coefficient at $\lambda_c$ is $0.4$ -- two times higher than our previous design above  -- which yields a $\sim 20\%$ reflectance; the percentage variation is $\sim 11\%$. The phase range in depletion is $\sim 64^{\circ}$, which is higher than what was reported in Fig. \ref{fig_g30_conn04} due to the fact that for $t_{pert}=2$ nm, the gap is almost completely depleted. However, the overall phase range ($\sim 305^{\circ}$) is lower than that reported in Fig. \ref{fig_g30_conn04}, which suggests that the phase range trades-off against reflectance. 

%case of $g=15$ nm, while the other optimized parameters of the dipole nanoantenna are $L_d=149$ nm, $a_x=298$ nm, and $p_c=0.5\cdot (L_d-w_c)$. 
%Designing the dipole nanoantenna in order to be able to fully deplete the gap appears as a promising strategy to increase the phase range in depletion. 
%This means that if we want to increase the phase range we have to accept a lower reflectance, and vice versa.

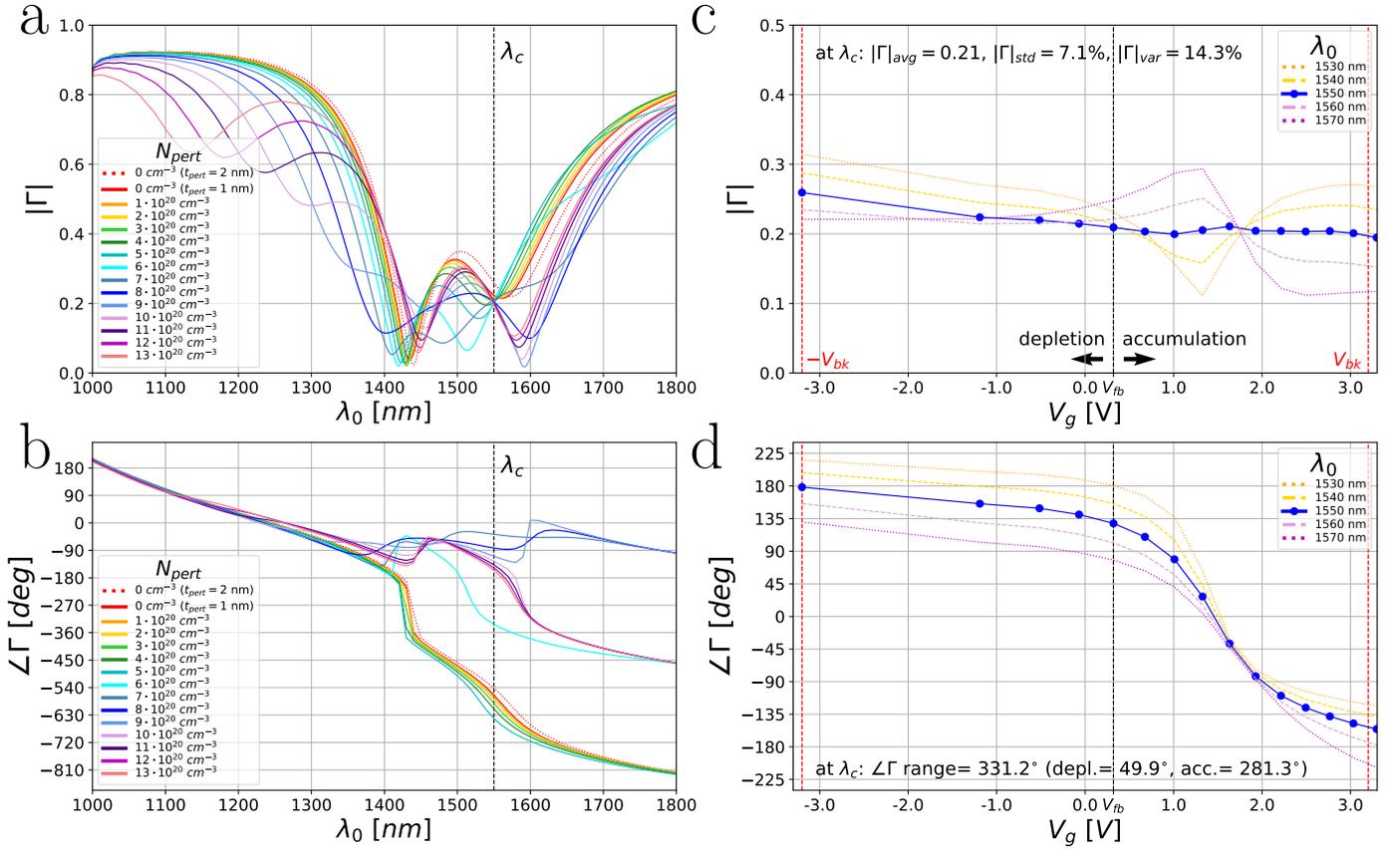
\begin{figure}[htbp]
\centering
\includegraphics[width=1\textwidth]{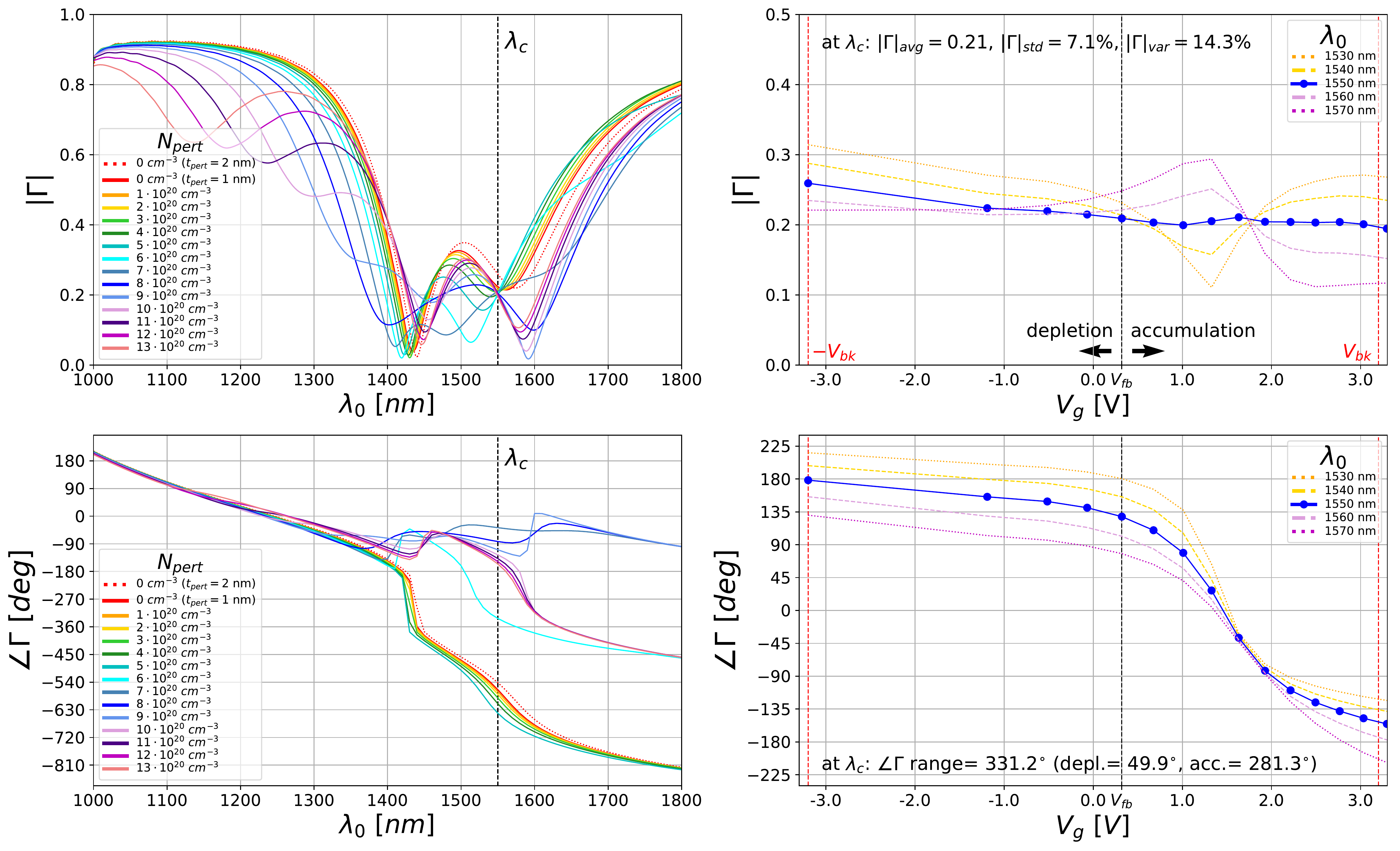} 
\caption{Third optimized pixel with parameters $g=30$ nm, $L_d=186$ nm, $a_x=404$ nm, $t_m=100$ nm, and $p_c=0.5\cdot (L_d-w_c)$: (a) $|\Gamma|$ and (b) $\angle \Gamma$ {\it vs} $\lambda_0$ for varying $N_{pert}$ within the $t_{pert}$ layer. (c) $|\Gamma|$ and (d) $\angle \Gamma$ {\it vs} $V_g$ for $\lambda_0$ near $\lambda_c$. The phase plots are unwrapped for clarity.}
\begin{picture}(0,0)
\put(-255,370){\Huge a} 
\put(0,370){\Huge c}
\put(-255,205){\Huge b}
\put(0,205){\Huge d}
\end{picture}
\label{fig_g30}
\end{figure}

In Fig. \ref{fig_g30}, we show the optical simulation results for a third pixel optimized to exhibit a constant amplitude point at $\lambda_c$. As an appendix to Fig. \ref{fig_g30}, we show in File 1 (Supporting Information) how the variation of the geometric parameters $a_x$, $t$, $t_m$, and $w$ affects the amplitude of the resonance dips of the $|\Gamma|$ curves ({\it i.e.}, they move  up and down), but with a negligible spectral shift. This affects the intersection of the curves and in turn the uniformity of the reflection coefficient at $\lambda_c$. In particular, we observe that decreasing $a_x$, or increasing $t$, $t_m$, or $w$ contribute to the formation of the constant amplitude point in a qualitatively similar way. Furthermore, by comparing Fig. \ref{fig_g30} to Fig. \ref{fig_g30_conn04}, we note that in Fig. \ref{fig_g30} we have slightly lower $|\Gamma|_{avg}$ and phase range, though $g=30$ nm in both cases. This suggests that merging the two resonances of the system, as was the case for the optimized design in Fig. \ref{fig_g30_conn04}, may provide a strategy to improve reflectance and phase range. This merging can be achieved by optimizing the connector position, to which we now turn.

% due to a non zero amplitude gradient. %This is achieved by optimizing the constant amplitude wavelength through the adjustment of the geometric parameters. 
%as ``tight'' as possible, {\it i.e.}, in a way that all the reflectivity curves cross in the same point. 
% and will be used to tune the simulation of the array to demonstrate beam steering. 
%%%%%%%%%%%%%%%%%%
%, which will be used to demonstrate beam steering in Section \ref{sec_array}.
% with $p_c=0.5\cdot (L_d-w_c)$, $L_d=186$ nm, $g=30$ nm, and pitch size $a_x=404$ nm  
%We consider an optimized pixel   %$w=50$ nm, $w_c=50$ nm, $t=50$ nm, $t_m=100$ nm, and $a_z=500$ nm.
%We use the same pixel used in Section \ref{sec_conn}, and . 
%This was used during the optimization of the designs proposed in this paper to make the magnitude of reflectivity as constant as possible. 
%, which can be seen in the context of parameters reduction for more efficient optimization. %, and this suggests that the existence of the crossing point is due to a balance between absorption in gold and ITO. 

%%%%%%%%% <-------------------
%which depends on the geometric parameters $L_d$, $g$, and $t_{ox}$ of the nanoantenna. 

\subsection{Effect of the connector position and dual-band pixel}\label{sec_conn}

We demonstrate in this subsection that the position of the connectors in the pixel significantly alters the optical response. This is an important aspect for the experimental realization of this type of structure. Here we consider only the case where the two sets of connectors are equally distant from the gap, {\it i.e.}, we use the same value of $p_c$ for both branches of the dipole nanoantenna. %; in general, each branch may have a connector at a different $p_c$, but we leave this for future research. 
For the optimized pixel reported in Fig. \ref{fig_g30_conn04}, we used $p_c=0.4\cdot (L_d-w_c)$, which results in the plasmonic system exhibiting only one resonance, rendering the connectors optically non-invasive. In the optimized pixels in subsection \ref{sec_g15}, we used $p_c=0.5\cdot (L_d-w_c)$, and this resulted in the reflection coefficient amplitude curves each exhibiting two resonances. Using the parameters of the pixel in Fig. \ref{fig_g30} but considering different connector positions, $p_c$, we will explore the effect of the connector location on the optical response, and the origin of these two resonances.

\begin{figure}[htbp]
\centering
\includegraphics[width=1\textwidth]{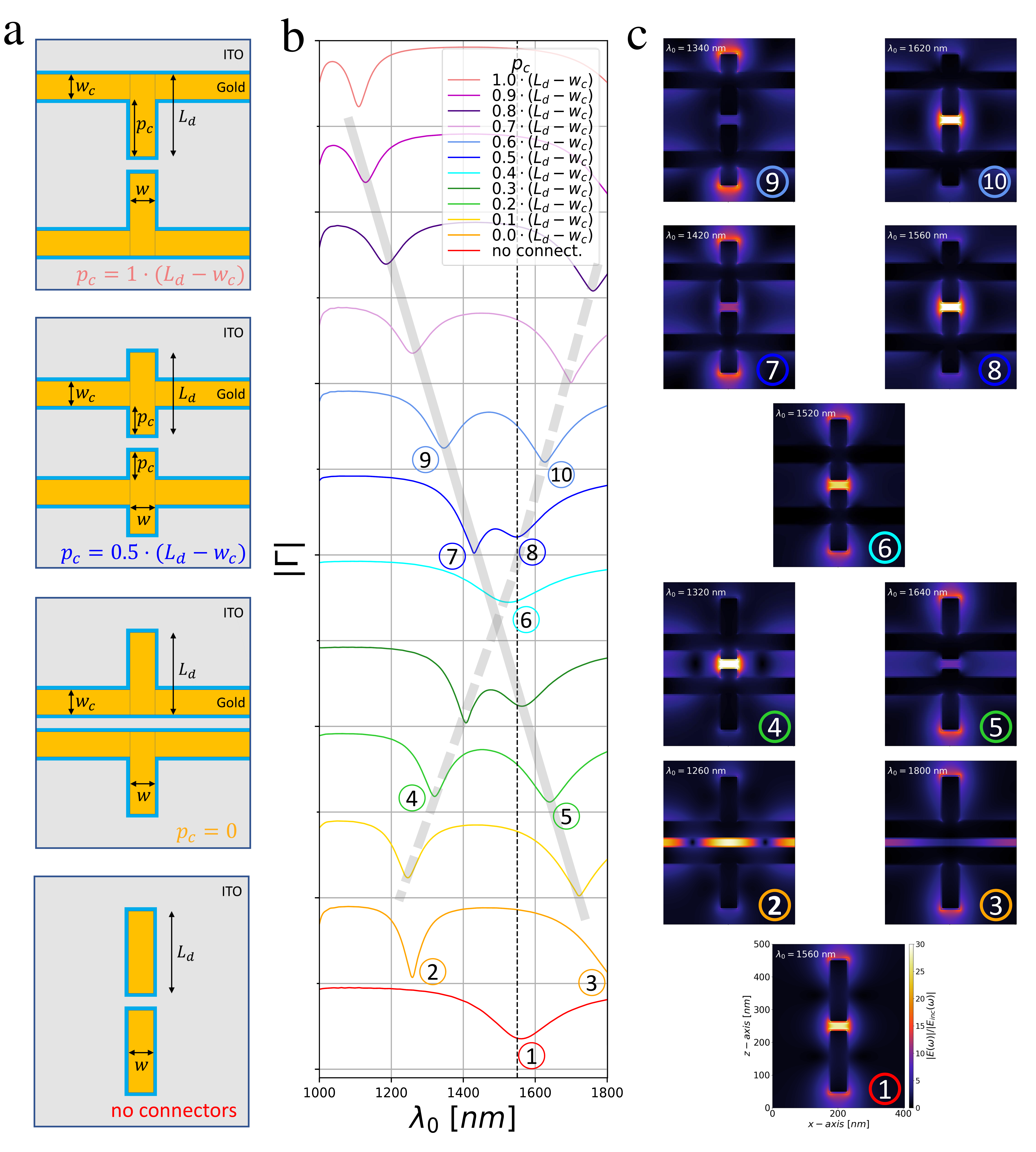}
\caption{Effect of the connector position: (a) illustration of different connector configurations, (b) reflection coefficient for varying connector positions as sketched in Fig. \ref{fig5}(a), (c) field distribution for some resonances highlighted in Fig. \ref{fig5}(b).}
\label{fig5}
\end{figure}

Considering here unperturbed ITO ($N_{pert}=N_0$), we varied the connector position from $p_c=0$ to $p_c=L_d-w_c$ to investigate how the resonances evolve with changing $p_c$. In Fig. \ref{fig5}(a), from bottom to top, we sketch a dipole nanoantenna without connectors, with connectors at position $p_c=0$ ({\it i.e.}, located on either size of the gap), with connectors at position $p_c=0.5\cdot(L_d-w_c)$ (middle of the branches), and with connectors at position $p_c=L_d-w_c$ (at the extremities of the dipole nanoantenna). In Fig. \ref{fig5}(b), we show the evolution of $|\Gamma|$ as a function of $p_c$.
For some of the resonances in the reflection coefficient, identified by numerals in Fig. \ref{fig5}(b), we plot the corresponding magnitude of the electric near-field in Fig. \ref{fig5}(c). We observe that the single resonance in the case with no connectors splits into two when connectors are introduced: the primary resonance produces field enhancement in the gap of the nanoantenna, whereas the secondary resonance produces field enhancement at the extremities of the branches of the dipole nanoantenna. In the no connector case (field distribution 1, bottom), field enhancement exists in both the gap and the branch extremities. As $p_c$ increases, the two resonances evolve in the spectrum tracing out an ``X'', {\it i.e.}, as the primary resonance red-shifts and the secondary resonance blue-shifts (see grey dashed line and grey solid line in Fig. \ref{fig5}(b), respectively). There is a connector position for which the two resonances overlap, which for our example is at $p_c=0.4\cdot(L_d-w_c)$ (field distribution identified by numeral 6). This mode strongly resembles the no connector case and identifies the position selected in order for the effect of the connectors on the resonance of the nanoantenna to be minimized. This is the connector position used in our optimized pixel in Fig. \ref{fig_g30_conn04}.

\begin{figure}[htbp]
\centering
\includegraphics[width=1\textwidth]{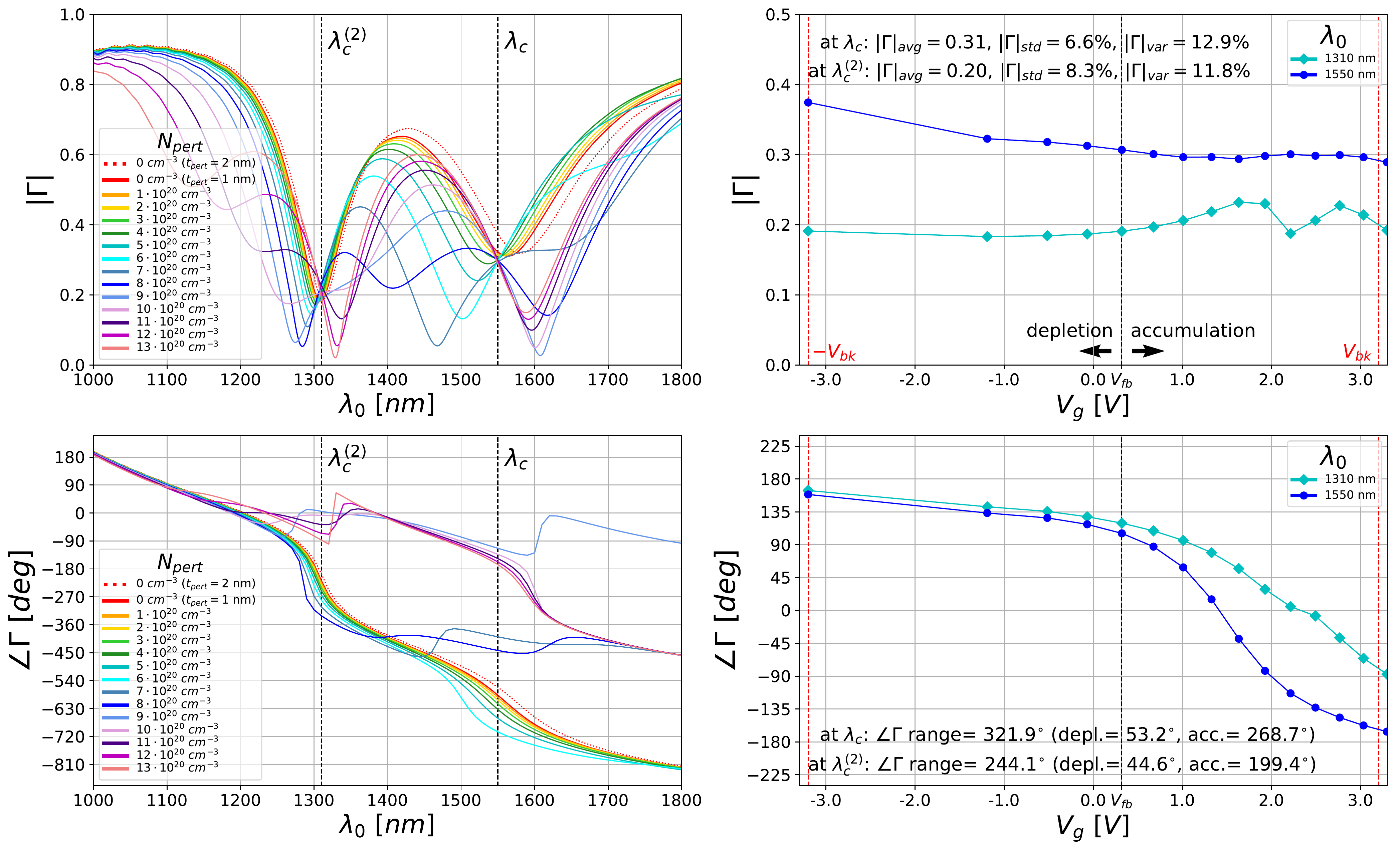} %_Ld-160_Ox3_x326_z500
\caption{Fourth optimized pixel with parameters with $g=20$ nm, $L_d=168$ nm, $a_x=348$ nm, $t_m=100$ nm, and $p_c=0.5\cdot(L_d-w_c)$: (a) $|\Gamma|$ and (b) $\angle \Gamma$ {\it vs} $\lambda_0$ for varying $N_{pert}$ within the $t_{pert}$ layer. (c) $|\Gamma|$ and (d) $\angle \Gamma$ {\it vs} $V_g$ for $\lambda_0$ near $\lambda_c$. The phase plots are unwrapped for clarity.}
\begin{picture}(0,0)
\put(-255,370){\Huge a} 
\put(0,370){\Huge c}
\put(-255,205){\Huge b}
\put(0,205){\Huge d}
\end{picture}
\label{fig_g20}
\end{figure}

However, the two resonances obtained by introducing the connectors can also be exploited.
In Fig. \ref{fig_g20}, we show results for a fourth optimized pixel that can be used for dual-band beam steering, as it exhibits two constant amplitude wavelengths at the telecom wavelengths $\lambda_c=1550$ nm and $\lambda_c^{(2)}=1310$ nm. The phase shifting at both operating wavelengths occurs as $\lambda_{ENZ}$ transitions through the resonance wavelengths.
At $\lambda_c$, we have the primary resonance with field enhancement in the gap. In this case, $|\Gamma|_{avg}\sim 0.31$ and the phase range is $\sim 321^{\circ}$, which follows the trend already observed in Figs. \ref{fig_g30_conn04} and \ref{fig_g15}, {\it i.e.}, reducing the gap size produces a higher magnitude of reflection coefficient and a lower phase range. 
At $\lambda_c^{(2)}$, we have the secondary resonance with field enhancement at the extremities of the dipole nanoantenna. The average magnitude of reflection coefficient is lower ($\sim 0.2$), as is the phase range ($\sim 244^{\circ}$). The lower phase range is due to the fact $\lambda_{ENZ}$ crosses $\lambda_c^{(2)}$ at higher voltages than it does for $\lambda_c$ such that the phase shifting mechanism described in subsection \ref{sec_ENZ} is not fully exploited. We would need to further increase $V_g$ ({\it i.e., $N_{pert}$}) to extend the phase range. 

\subsection{Electrostatic simulations}\label{sec_DEVICE}
In this subsection, we use electrostatic modelling to justify the choice of the perturbation parameters $N_{pert}$ and $t_{pert}$ used to simulate in the previous subsections the optical response of our pixels due to voltage biasing. We first consider a 1D cross-sectional model of our MOS capacitor, and then we present a 2D model which takes into account corner configurations which can be found in nanostructures.

\subsubsection{Simulation details}
\begin{figure}[htbp]
\centering
\includegraphics[width=1\textwidth]{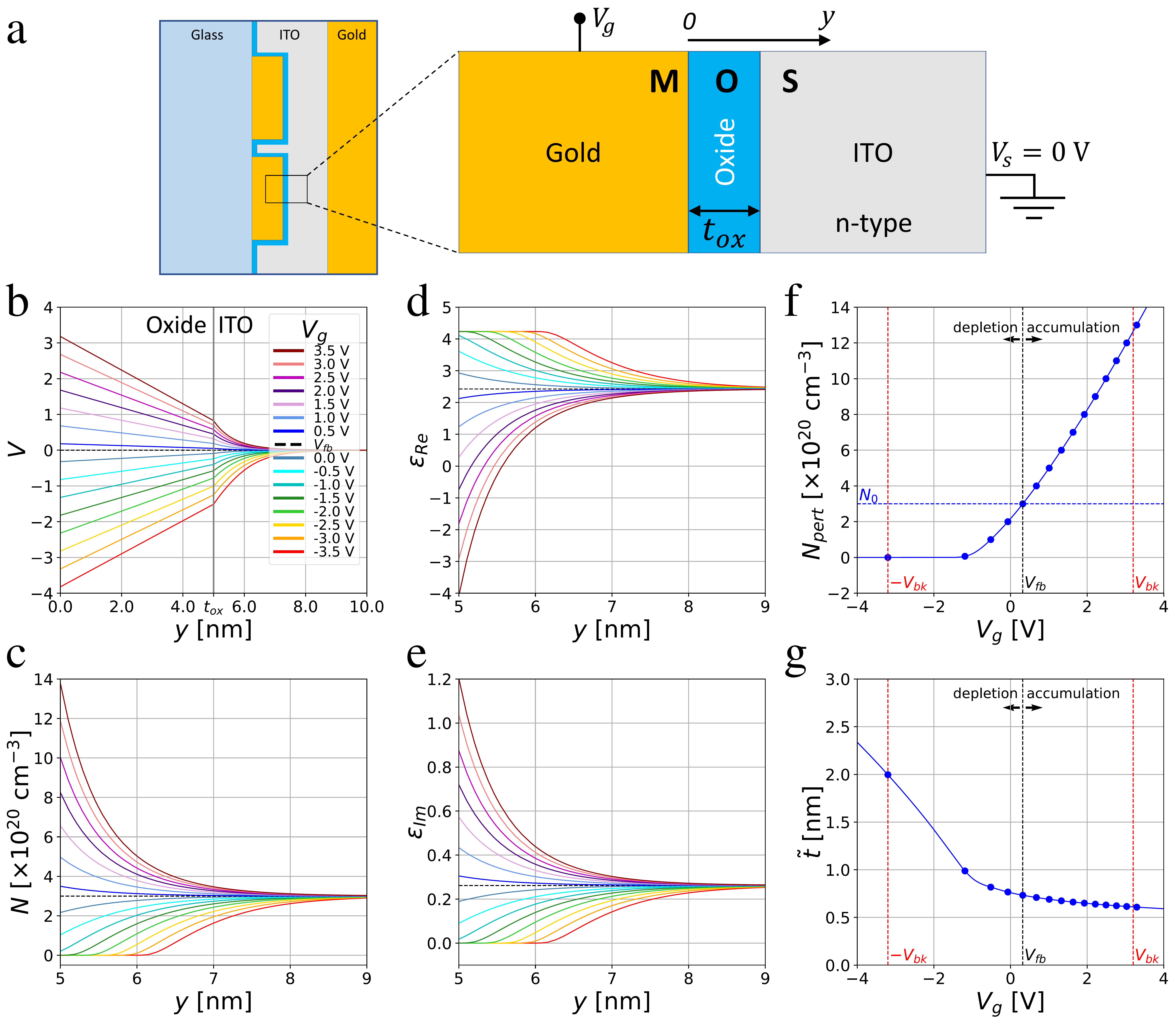}
\caption{(a) Side view of the plasmonic pixel and zoom of the MOS 1D cross-section modelled. (b) Voltage across the oxide and ITO for varying $V_g$. (c) Carrier density across ITO for varying $V_g$. (d, e) $\varepsilon_{Re}$ and $\varepsilon_{Im}$ across ITO at $\lambda_c$ for varying $V_g$. (f) $N_{pert}$ as a function of $V_g$. (g) Effective perturbed ITO thickness $\tilde t$ as a function of $V_g$. Figs. \ref{fig2}(b)--(e) share the same legend.}
\label{fig2}
\end{figure}

The electrostatic modelling was conducted using Lumerical DEVICE, which solves the Poisson and drift-diffusion equations self-consistently \cite{Lumerical}. We assume a work function for gold of $\phi_m=5.1$ eV. For the oxide, we consider \ce{HfO2}, which is a high dielectric constant material of static relative permittivity $\varepsilon_{DC}^{(\ce{HfO2})}=25$ \cite{Wilk2001,Robertson2006}.
ITO is treated as an n-doped semiconductor with $\varepsilon_{DC}^{(ITO)}=9.3$, electron affinity $\chi_s=4.8$ eV, band gap energy $E_g=2.8$ eV, effective mass of electrons as in optical simulation ($m_n^*=0.35\cdot m_e$), and effective mass of holes $m_p^*=m_e$ \cite{KafaieShirmanesh2018}. Since $E_g$ is large, the intrinsic carrier concentration $n_i$ is very small. We assume an unperturbed carrier (electron) density in ITO of $N_0=3\cdot 10^{20}$ cm$^{-3}$ \cite{KafaieShirmanesh2018}, which corresponds to a Fermi energy located within the conduction band, so that ITO behaves as a degenerate semiconductor.
We consider ITO at a potential of $V_s=0$ V (grounded) and we apply a gate voltage $V_g$ to the metal as depicted in Fig. \ref{fig2}(a).

Breakdown fields for oxides used in MOS systems depend strongly on the deposition process. Assuming a breakdown field for \ce{HfO2} of $E_{bk}=6.4$ MV/cm \cite{Olivieri2015}, for a nominal  thickness of $t_{ox}=5$ nm, we obtain a breakdown voltage of $V_{bk}=3.2$ V. The parameters of \ce{HfO2} are similar to those of HAOL, {\it i.e.}, a nanolaminate comprised of layers of \ce{Al2O3} and \ce{HfO2} ($\varepsilon_{DC}^{(\ce{HAOL})}=22$ and $E_{bk}=7.2$ MV/cm \cite{KafaieShirmanesh2018}), which could also be used in our phase shifting pixel as the insulating layer.

\subsubsection{1D cross-section model}
In Fig. \ref{fig2}(b), we show the potential $V$ across the structure (oxide+ITO) for varying $V_g$, which reveals a flat band voltage $V_{fb}=0.32$ V (value of $V_g$ at which $V(y)=0$). This means that for $V_{g}>V_{fb}$ the ITO layer goes into accumulation, whereas for $V_{g}<V_{fb}$ we have depletion.

 %We observe that, for equal $V_g$, the potential drop across \ce{HfO2} is lower than the potential drop across \ce{Al2O3}. This causes a larger excursion of the potential in ITO for \ce{HfO2} and thus a larger perturbation in the carrier density $N$, as shown in Figs. \ref{fig2}(d) and \ref{fig2}(e). %In \ref{fig2}(c) and \ref{fig2}(d), $\Delta V$ represent the potential difference across the stack, where $V_s=0$ V is used as a reference. 
%The range of $V_g$ is consistent with the breakdown voltage of HAOL and Hafnia oxides with thickness $t_{ox}=5$ nm. %In Figs. \ref{fig2}(d) and \ref{fig2}(e), we show the concentration of electrons $N$ across ITO for varying $V_g$ and for \ce{Al2O3} and \ce{HfO2}, respectively. These graphs show that the perturbation on the carrier density with respect to the nominal value $N_0$ increases for increasing $V_g$. 
%From Figs. \ref{fig2}(b) and \ref{fig2}(c), we note that 
%Antonio: should we add the case Vg=0.32V explicitly to  he graphs?

From Fig. \ref{fig2}(c), we note that the carrier density in ITO increases with increasing gate voltage $V_g$. In order to maximize this carrier perturbation, we need to use an oxide with a high product of $\varepsilon_{DC}\cdot E_{bk}$.
When no voltage is applied ($V_g=0$ V), the MOS capacitor is slightly in depletion; we need to apply $V_g=V_{fb}$ in order to have an unperturbed carrier density across ITO, {\it i.e.}, $N_{pert}=N_0$ within the $t_{pert}$ layer.

In Figs. \ref{fig2}(d) and \ref{fig2}(e), respectively, we show the complex permittivity $\varepsilon_{Re}$ and $\varepsilon_{Im}$ (at $\lambda_c=1550$ nm, {\it i.e.}, the wavelength at which our pixels were optimized) by varying $V_g$ and then $N_{pert}$ within the perturbed ITO layer.
We observe that for increasing $V_g$, $\varepsilon_{Re}$ decreases becoming negative, and $\varepsilon_{Im}$ increases, {\it i.e.}, the perturbed ITO layer becomes more and more metallic. 
Fig. \ref{fig2}(c) also shows that the perturbation of the carrier density is localized within a thickness of $\sim 1-2$ nm. In order to accurately resolve this perturbation, electrostatic computations were conducted with a discretization of $\sim0.05$ nm, but this space step is challenging to apply in the corresponding 3D optical simulations, which are carried out here using a discretized space step of 1 nm. 
To model the perturbed carrier density in optical simulations, we need to homogenize the perturbation of the carrier density in ITO with respect to the unperturbed level $N_0$. This is done by considering the $N(y)$ curves in Fig. \ref{fig2}(c), and by identifying a constant carrier density $\tilde N$ and a constant thickness $\tilde t$ such that 
\begin{equation}
\tilde N \cdot \tilde t= \int^{+\infty}_{t_{ox}}(N(y)-N_0)dy. %\tilde N = (N_{pert}-N_0)
\label{eq_DEVICE}
\end{equation}
We set $\tilde N=\lim_{y \to t_{ox}^+} N(y)-N_0$, that is, we use the limit value that $N(y)$ reaches at the boundary between the oxide and ITO. Finally, we retrieve $N_{pert}=\tilde N+N_0$, as plotted in Fig. \ref{fig2}(f) for varying $V_g$ within the breakdown voltage range (identified by vertical red dashed lines). We note that $N_{pert}$ varies between $0$ (depletion regime) and $\sim13\cdot10^{20}$ cm$^{-3}$ (accumulation regime) as used in the optical simulations. Inverting Eq. (\ref{eq_DEVICE}), we derive $\tilde t$, which is shown in Fig. \ref{fig2}(g) as a function of $V_g$. In accumulation, we obtain a nearly constant $\tilde t$ slightly lower than 1 nm, that we approximate with a perturbed ITO layer $t_{pert}=1$ nm in the optical simulations. In depletion, the value of $\tilde t$ increases up to $\sim 2$ nm with decreasing $V_g$. The case of full depletion with $t_{pert}=2$ nm was exploited in the optical simulations to further extend the phase range of the pixel. The blue dots in Figs. \ref{fig2}(f) and \ref{fig2}(g) correspond to the blue dots in the optical simulation results for the four optimized pixels reported in Figs. \ref{fig_g30_conn04}, \ref{fig_g15}, \ref{fig_g30}, and \ref{fig_g20}.

As the $t_{pert}=1$ nm approximation in the accumulation regime may overestimate the pixel performance, we performed optical simulations where the thickness of the perturbed layer was taken to be $t_{pert}=0.5$ nm, and the simulation domain was discretized with a space step of 0.5 nm. In this case, by optimizing the pixel to find a constant amplitude point at $\lambda_c$, we obtained a lower magnitude of the reflection coefficient and an increased phase range, following the trade-off between phase range and amplitude. 
This result suggests that the phase range of the reflection coefficient at $\lambda_c$ depends on the range of $N_{pert}$, while its amplitude depends on the thickness $t_{pert}$ of the perturbed ITO layer.

\subsubsection{2D correction to 1D model}

\begin{figure}[htbp]
\centering
\includegraphics[width=1\textwidth]{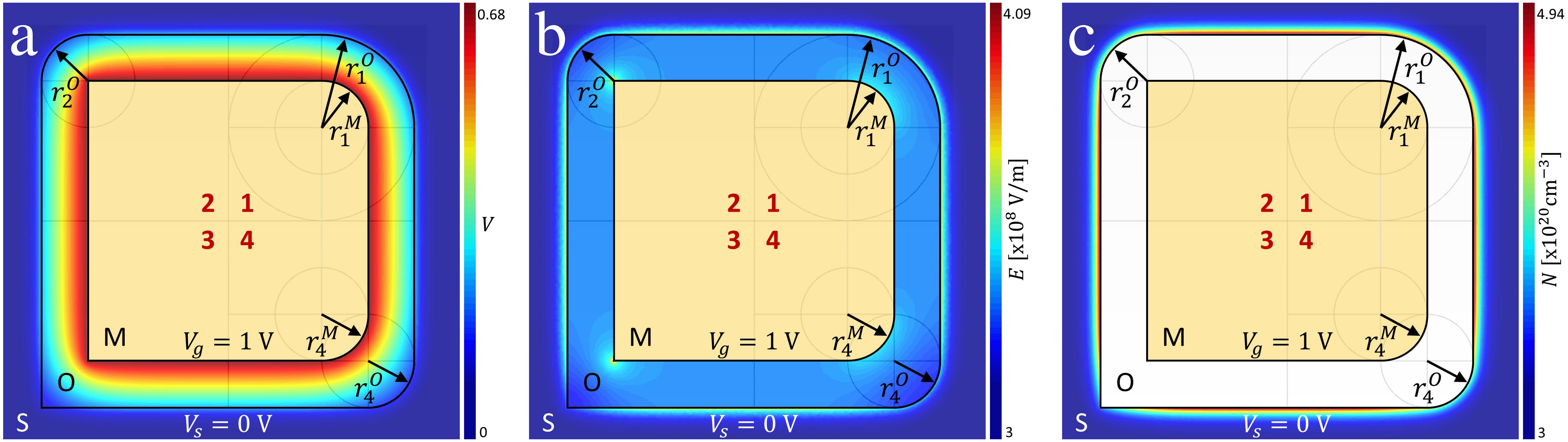} %_Lm-110_Ox3_x290_z500
\caption{(a) Potential, (b) electric field, and (c) carrier density distributions from electrostatic simulation of a representative MOS structure with four types of rounded corners for $V_g = 1$ V.}
\label{fig3}
\end{figure}

The MOS structure analyzed in 1D in Fig. \ref{fig2} does not take into account the geometry of a nanoantenna. Since our plasmonic pixel (Fig. \ref{fig1}) is more complex as corners are involved, we performed a MOS study for a 2D geometry containing four different corner types as shown in Fig \ref{fig3}. In quadrant 1 (top right), we consider a rounded metal corner of radius $r_1^M=5$ nm and a rounded oxide corner of radius $r_1^O=r_1^M+t_{ox}=10$ nm. In quadrant 2 (top left), we have a square metal corner and a rounded oxide corner with $r_2^O=5$ nm. In quadrant 3 (bottom left), both metal and oxide corners are square. In quadrant 4, both metal and oxide corners are rounded and of the same radius $r_4^M=r_4^O=5$ nm. Figs. \ref{fig3}(a), \ref{fig3}(b) and \ref{fig3}(c) show, respectively, the potential distribution $V(x,z)$, the electric field distribution $|E(x,z)|$ and the carrier density distribution $N(x,z)$ for $V_g=1$ V and ITO grounded ($V_s=0$ V).
We observe that $V$, $E$, and $N$ exhibit a different behaviour at the corners compared to regions away from the corners where the behaviour of the 1D system is recovered. As the corner becomes more rounded (quadrant 1 in Fig. \ref{fig3}(c)), the carrier density $N$ approaches that calculated for the 1D system. These results suggest that the perturbed layer in ITO should be modelled by a variable thickness depending on the roundness of the corners. However, the results of optical simulations of a nanoantenna with corners as in quadrant 1 (not shown), reproduce the physical operation of pixels containing all square corners with uniform $t_{pert}$; thus, for ease of simulation, square corners and uniform $t_{pert}$ only are used throughout this paper. To consider round corners, the pixels as presented here would need to be re-optimized, as round corners produce a blue shift in the optical response due to the smaller total volume of rounded nanostructures versus square ones, all other geometrical parameters being equal.

%\subsection{Methods}\label{sec_methods}
%\subsubsection{Optical simulation parameters}
%\subsubsection{Electrostatic simulation parameters}

%%%%%%%%%%%%%%%%%%%%%%%

\section{Simulation of phased arrays for 1D beam steering}\label{sec_array}

\begin{figure}[htbp]
\centering
\includegraphics[width=1\textwidth]{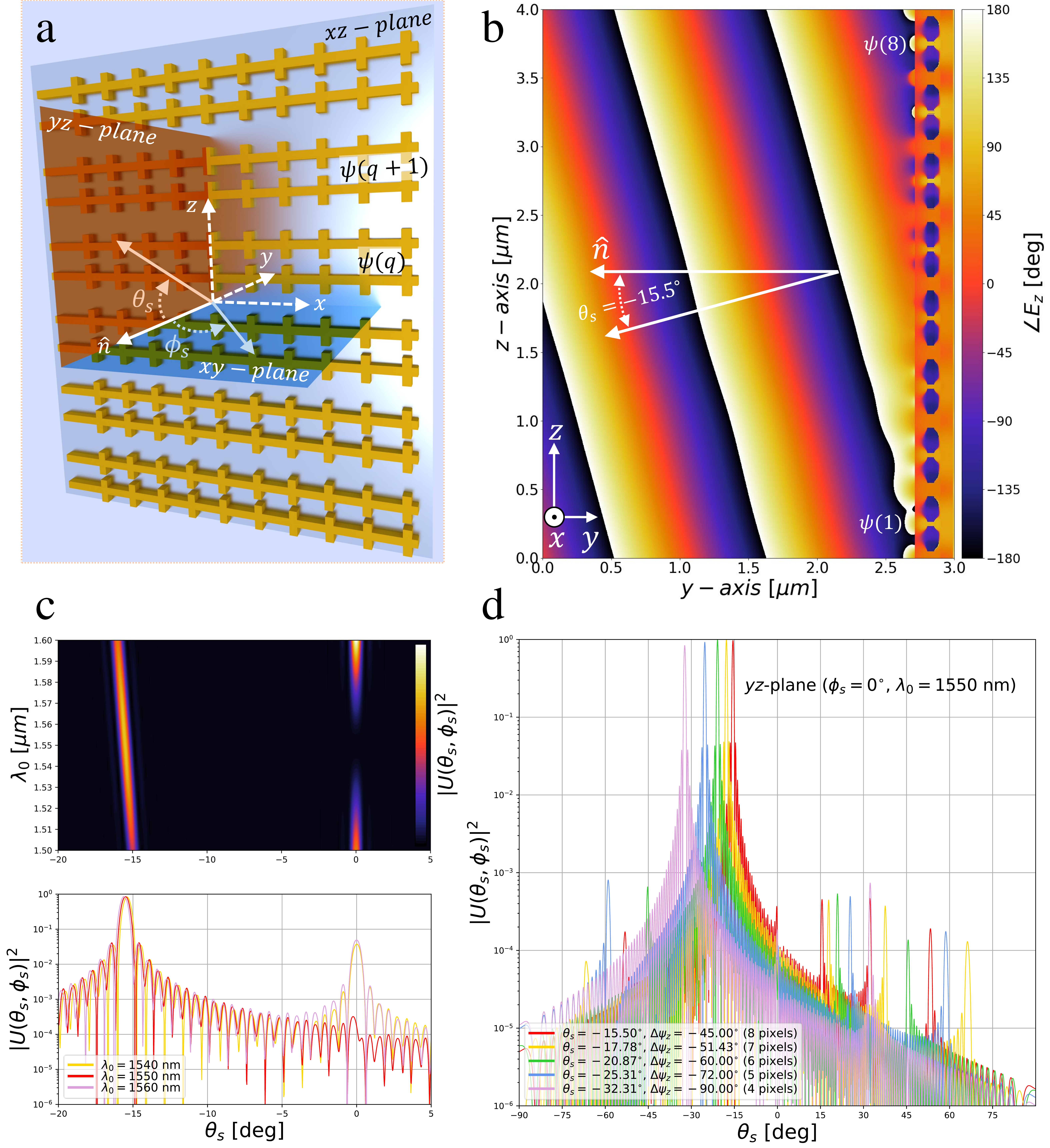}
\caption{(a) Sketch of an array of pixels with coordinates, (b) $\angle E_z$ and (c) radiation pattern in the $yz$-plane ($\phi_s=0^{\circ}$) for an extended unit cell ($M_z=8$, $\theta_s=-15.5^{\circ}$), (d) radiation pattern for extended unit cells with $M_z=4,5,6,7,8$ pixels.}
\label{fig6}
\end{figure}

In Section \ref{sec_pixel}, we designed our pixels such that the pitch is small enough to avoid grating lobes due to the pitch size ($a_z, a_x<\lambda/2$) for any steering angle, yet large enough to avoid coupling between adjacent pixels due to near-field interaction. 
In order to validate our pixel designs for beam steering, we numerically test one of them (third optimized pixel in Fig. \ref{fig_g30}) in an array configuration. 
While in Section \ref{sec_pixel} we optimized the response of our pixels by applying periodic boundary conditions, which is equivalent to simulating an infinite array of pixels under the same perturbation, here, we create an array of pixels in the $xz$-plane, as sketched in Fig. \ref{fig6}(a). The arrangement of connectors makes the system controllable row by row, which means that all nanoantennas in the same row will share the same potential, and thus the same phase of the reflection coefficient. By varying the phase associated with the pixels along $z$, we produce beam steering in the $yz$-plane ($\phi_s=0$) with a steering angle $\theta_s$; in our example there is no steering in the $xy$-plane.

We cannot simulate arbitrarily large arrays with our FDTD model, so we apply periodic boundary conditions on an extended unit cell containing one pixel along $x$ and $M_z$ pixels along $z$. Thus, along $z$, each pixel produces a different phase of the reflection coefficient $\psi(q)$, with the position of the pixels in $z$ labelled by $q=1, ..., M_z$. To enforce a phase gradient along $z$ in the array, which is necessary for beam steering, we need a constant phase difference between adjacent pixels along $z$, $\Delta\psi_z=\psi(q+1)-\psi(q)$. This sets a condition on $\Delta\psi_z$ in our simulations, which can only assume the values $|\Delta\psi_z|=2\pi/M_z$. 
Knowing $\Delta\psi_z$, the steering angle $\theta_s$ can be analytically determined via the generalized form of Snell's law of reflection \cite{Yu2011}:
\begin{equation}
k\sin\theta_s=\frac{\Delta\psi_z}{a_z}+ k\sin\theta_i,
\label{Snell}
\end{equation}
where $k=2\pi n_{SiO_2}/\lambda_c$, $\theta_i$ is the angle of incidence (in our case, we have normal incidence, {\it i.e.}, $\theta_i=0$) and $\theta_s$ is the angle of the reflected/steered beam.

We consider an extended unit cell with $M_z=8$ and $\Delta\psi_z=-45^{\circ}$. Using the curve for $\lambda_c=1550$ nm in Fig. \ref{fig_g30}(d), we determine the voltage that must be applied to each row to produce the required phase of the reflection coefficient $\psi(q)$ for each nanoantenna in the extended unit cell. Note that applying a voltage in an optical simulation corresponds to using a specific value of $N_{pert}$ derived via the relation between $V_g$ and $N_{pert}$ in Fig. \ref{fig2}(f). The array is excited by a monochromatic plane wave at $\lambda_c$ which is $z$-polarized and normally incident on the metasurface. The reflected field is obtained by subtracting the incident field from the total field.
In Fig. \ref{fig6}(b), we plot the phase of the reflected field $E_z$, from which we note that the phase front of the reflected wave is flat and propagates at an angle of  $\theta_s=-15.5^{\circ}$ with respect to the normal to the plane of the array, as confirmed by Eq. (\ref{Snell}). %In the region close to the metasurface, we observe near-field wiggles, which rapidly decay with distance from the metasurface. 
We find that the amplitude of the reflected field $E_z$ is nearly uniform with a percentage variation $|E_z|_{var}=5.6\%$. These phase and amplitude profiles indicate beam steering.

The flat phase front reflected from the array only provides qualitative information about the steering. To further validate the beam steering, the radiation pattern in the far-field must be observed. From the near-fields in the reflection region, we calculate the far-fields via a near-to-far-field transformation \cite{Taflove2005}, and then obtain the radiation pattern by squaring the far-fields. This near-to-far-field transformation is carried out considering $P_z=25$ periods of the extended unit cell such to mimic an array with 200 pixels. The radiation pattern allows us to quantify how the electromagnetic power is distributed in space as a function of angle, and thus, to quantify the quality of the beam steering in terms of intensity of the long-period grating lobes. In Fig. \ref{fig6}(c), we show the normalized radiation pattern $|U(\theta_s,\phi_s)|^2$ in the $yz$-plane ($\phi_s=0^{\circ}$). In the top panel we show the contour plot of the far-field reflected power as a function of wavelength (varying from $\lambda_0=1500$ nm to $1600$ nm) and far-field angle $\theta_s$. The contour plot reveals that near $\lambda_c$ the specular reflection in the normal direction disappears (black region in the colour band for $\theta_s=0^{\circ}$). Looking at this black region, we calculate the wavelength range such that the far-field intensity in the normal direction is two orders of magnitude lower than the main beam, corresponding to a beam steering bandwidth of $\sim 10$ nm. The bottom panel contains three curves extracted from the contour plot on a log scale, where we see that for $\lambda_c=1550$ nm the long-period grating lobe at $\theta_s=0^{\circ}$ is more than two orders of magnitude less intense than for $\lambda_0=1540$ and 1560 nm. These results validate the optimization at $\lambda_c$ of the single pixel, as they show that at $\lambda_c$ the specular reflection is negligible and the beam is completely steered. %The bottom panel gives the intensity of the far-field on a log scale, showing .

In Fig. \ref{fig6}(d), we show the radiation patterns for different values of $M_z=\{4,5,6,7,8\}$, calculated via near-to-far-field transformation using $P_z= \{50,40,34,29,25\}$, respectively, such that $M_z\cdot P_z\sim 200$. These $M_z$ values correspond to decreasing values of $\Delta\psi_z$ and thus steering angle $\theta_s$, as reported in the legend of Fig. \ref{fig6}(d). From phased antenna array theory, we know that the radiation pattern of the array is obtained by multiplying the radiation pattern of the single emitter $F$ with an array factor $A$ that depends on the arrangement of the emitters, {\it i.e.}, $|U(\theta_s,\phi_s)|^2=|F(\theta_s,\phi_s)|^2|A(\theta_s,\phi_s)|^2$. As $A(\theta_s,\phi_s)$ reaches the same maximum value for any steering angle, the fact that the peak of the radiation patterns in Fig. \ref{fig6}(d) decreases with increasing steering angle suggests that each nanoantenna exhibits a dipole-like radiation pattern in reflectance. All the radiation patterns show long-period grating lobes which are more than three orders of magnitude lower than the main lobe, confirming beam steering over a wide angular range. We note that this array simulation approach does not take into account the limited phase range of the pixel, as $M_z$ is chosen such that $(M_z-1)|\Delta\psi_z|$ is lower than the phase range of the pixel. The long-period grating lobes that we observe are only due to the non-uniform amplitude of the reflection coefficient as $V_g$ varies, and they obey the diffraction grating equation:
\begin{equation}
\sin\theta_s^{(m)}=m\lambda/(a_zM_z),
\end{equation}
where $a_zM_z$ is the size of the extended unit cell (our long period), and $m$ is the diffraction order ($m=1$ identifies the main lobe).
Movies 1 and 2 (Supporting Information) show the time-domain evolution of the absolute value of the electric field for $M_z=8$ ($\Delta\psi_z=-45^{\circ}$, $\theta_s=-15.5^{\circ}$) and $M_z=4$ ($\Delta\psi_z=-90^{\circ}$, $\theta_s=-32.31^{\circ}$). The scattered-field region contains only the steered beam and is thus made large in the simulation for visual clarity, while the total-field region contains both the reflected and incident fields, and is thus limited to the vicinity of the metasurface where the input signal is injected. As this is a time-domain simulation, we find that the steering does not occur instantaneously, as the phase relations responsible for the creation of the steered beam take some time to build up.

\section{Conclusion}
We propose a novel plasmonic pixel for beam steering applications in reflectarray configurations. The pixel is of sub-wavelength dimensions, and exhibits a reflection coefficient that, at a specific operating wavelength, varies with applied voltage over a large phase range with nearly constant amplitude. The structure consists of a plasmonic dipole nanoantenna on a substrate covered conformally by a thin oxide layer and ITO. Electrically, the structure forms a MOS capacitor with the nanoantenna acting as the electrical contact of the capacitor. The MOS capacitor is exploited to perturb the free electron density within a region of ITO in contact with the oxide. This perturbed layer of ITO will experience a change of its epsilon-near-zero wavelength as a function of applied voltage, eventually crossing the resonance wavelength of the nanoantenna.
When this occurs, the environment surrounding the nanoantenna switches from all dielectric to one containing a metallic shell. This abruptly changes the resonance of the nanoantenna and causes a large phase shift in its reflection coefficient. 

We found that the phase range of the reflection coefficient depends on the maximum carrier density perturbation induced in ITO which is bounded by the oxide breakdown voltage. Moreover, the magnitude of the reflection coefficient at the operating wavelength depends on the thickness of this perturbation, but can be made uniform across the phase range by optimizing the geometrical parameters. We analyzed the performance of several pixel designs and found that a phase range of $330^{\circ}$ is possible with a nearly flat magnitude of the reflection coefficient of 0.2; higher magnitudes of $\sim 0.4$ are possible if we can accept a phase range of $300^{\circ}$. 

We also found that there is an optimal position for the electrical connectors where they contribute minimally to the optical response of the nanoantenna. However, in an alternate position, their effect may be exploited for realizing dual-band beam steering. Though the performance of the proposed pixel is primarily limited by the breakdown field of the oxide, the phase and amplitude responses of the reflection coefficient achieved with materials currently available are highly promising for beam steering applications, as demonstrated by our 3D optical simulations of  beam steering in 1D.

%a large phase range and a nearly flat amplitude of the reflection coefficient, which are all desired characteristics in optical phased arrays for LIDAR beam steering to minimize the amplitude of the grating lobes. 
%We analyzed the performance of several configurations in terms of the magnitude and phase response of the reflection coefficient, and found that a phase range of $\sim 330^{\circ}$ is possible with a nearly flat magnitude of the reflection coefficient in the range of $\sim 0.2$. The phase range and the magnitude of the reflection coefficient are trade-offs: the latter becomes $\sim 0.4$ if we can accept a phase range of $\sim 300^{\circ}$. 
%The phase and magnitude of the reflection coefficient achieve in this study are very promising to realize beam steering in LIDAR systems, but more research is needed to further maximize the magnitude for power efficiency. Different gating of the two branches of the dipole nanoantenna can also be exploited to improve the performance of the plasmonic pixel.

%\section*{Funding}
%SOSCIP, NSERC, CRC

\section*{Acknowledgments}
We acknowledge computational support from Compute Canada, and financial support from NSERC and Huawei Technologies Canada.

% Bibliography
\bibliography{Lidar,extra}

\end{document}